\documentclass[english,onecolumn]{IEEEtran}
\usepackage[T1]{fontenc}
\usepackage[latin9]{inputenc}
\usepackage{babel}
\usepackage{float}
\usepackage{amsmath}
\usepackage{amsthm}
\usepackage{amssymb}
\usepackage{esint}
\usepackage[unicode=true]
 {hyperref}

\makeatletter

\floatstyle{ruled}
\newfloat{algorithm}{tbp}{loa}
\providecommand{\algorithmname}{Algorithm}
\floatname{algorithm}{\protect\algorithmname}

\theoremstyle{plain}
\newtheorem{thm}{\protect\theoremname}
\theoremstyle{remark}
\newtheorem{rem}[thm]{\protect\remarkname}
\theoremstyle{definition}
\newtheorem{example}[thm]{\protect\examplename}
\theoremstyle{plain}
\newtheorem{prop}[thm]{\protect\propositionname}
\theoremstyle{definition}
\newtheorem{defn}[thm]{\protect\definitionname}

\usepackage{mathrsfs}
\usepackage{algorithm,algpseudocode}

\usepackage{colortbl}
\definecolor{lightgray}{rgb}{0.9,0.9,0.9}
\definecolor{lightred}{rgb}{1,0.8,0.8}
\definecolor{lightgreen}{rgb}{0.6,1,0.6}
\definecolor{lightyellow}{rgb}{1,1,0.5}
\definecolor{lightgrey}{rgb}{0.8,0.8,0.8}

\allowdisplaybreaks[1]

\makeatother

\usepackage{listings}
\providecommand{\definitionname}{Definition}
\providecommand{\examplename}{Example}
\providecommand{\propositionname}{Proposition}
\providecommand{\remarkname}{Remark}
\providecommand{\theoremname}{Theorem}

\begin{document}
\title{An Automated Theorem Proving Framework for Information-Theoretic Results}
\author{Cheuk Ting Li\\
Department of Information Engineering, The Chinese University of Hong
Kong\\
Email: ctli@ie.cuhk.edu.hk\thanks{This paper was presented in part
at the IEEE International Symposium on Information Theory 2021. A
brief description of the contents of the conference version was also
included in one subsection in an article in IEEE BITS the Information
Theory Magazine.}}
\maketitle
\begin{abstract}
We present a versatile automated theorem proving framework capable
of automated discovery, simplification and proofs of inner and outer
bounds in network information theory, deduction of properties of information-theoretic
quantities (e.g. Wyner and G{\'a}cs-K{\"o}rner common information),
and discovery of non-Shannon-type inequalities, under a unified framework.
Our implementation successfully generated proofs for 32 out of 56
theorems in Chapters 1-14 of the book \emph{Network Information Theory}
by El Gamal and Kim. Our framework is based on the concept of existential
information inequalities, which provides an axiomatic framework for
a wide range of problems in information theory.
\end{abstract}


\section{Introduction}

In information theory, information is commonly represented as random
variables. Proofs of theorems in information theory often rely on
properties of random variables, e.g. implications between conditional
independence statements \cite{dawid1979conditional,spohn1980stochastic},
and inequalities among entropy and mutual information terms. Computer
programs for automated deduction of these properties are therefore
useful for proving information-theoretic results. For example, the
linear programming approach to proving linear information inequalities
studied by Yeung \cite{yeung1997framework} and Zhang and Yeung \cite{zhang1997non,zhang1998characterization}
(implemented in the Information Theoretic Inequality Prover \cite{yeung1996itip})
has found success in proving results on network coding \cite{yeung2008information,chan2008dualities,yan2012implicit,apte2014algorithms,apte2016explicit}
and secret sharing \cite{metcalf2011improved,farras2018improving,gurpinar2019use}.

In this paper, which is the complete version of \cite{li2021automatedisit}\footnote{The short version \cite{li2021automatedisit} only includes the definition
of existential information inequalities and the auxiliary searching
algorithm. It does not include other results in this paper (e.g. inner
and outer bounds in network information theory, simplification of
regions, etc.). A very brief description of existential information
inequalities and the PSITIP program was also given in one subsection
in an article in IEEE BITS the Information Theory Magazine \cite{yeung2021machine}.
Note that \cite{yeung2021machine} does not include any of the technical
results in this paper.}, we develop a versatile automated theorem proving framework capable
of automated discovery, simplification and proofs of inner and outer
bounds in network information theory involving auxiliary random variables,
deduction of properties of information-theoretic quantities, and discovery
of non-Shannon-type inequalities. To this end, we study the inference
problem of existential information inequalities, which is a generalization
of the inference problem of linear information inequality. Existential
information inequalities concern the problem of deciding whether there
exist some random variables satisfying certain constraints on their
entropy and mutual information terms. For example, the copy lemma
\cite{zhang1998characterization,dougherty2011non}, which states that
for any random variables $X,Y$, there exists a random variable $U$
that is conditionally independent of $Y$ given $X$, such that the
joint distribution of $(X,U)$ is the same as that of $(X,Y)$, can
be regarded as an existential information inequality. Therefore, the
technique for proving non-Shannon-type inequalities using the copy
lemma can be included in the framework of existential information
inequalities.

Existential information inequalities are especially suitable for network
information theory, where the problem of finding auxiliary random
variables in proving converse results can be expressed as existential
information inequalities. Refer to \cite{elgamal2011network} for
various examples. For achievability results, while general coding
theorems have been studied, for example, in \cite{rini2013general,minero2015unified,lee2015unified,lee2018unified},
the inner bounds produced by these methods are often more complicated
than necessary, and manual manipulations are needed to simplify these
bounds to more familiar forms. The simplification of rate regions
can also be performed using existential information inequalities.
The algorithm in this paper is capable of automatically computing
inner and outer bounds for a multiuser setting, requiring only a simple
description of the communication network as the input to the program.
Examples of results provable by this framework include channels with
state (Gelfand-Pinsker theorem \cite{gelfand1980coding}, achievability
and converse), lossy source coding with side information (Wyner-Ziv
theorem \cite{wyner1976ratedistort}, achievability and converse),
multiple access channel \cite{ahlswede1971multi,liao1972multiple,ahlswede1974capacity}
(achievability and converse), broadcast channel (Marton's inner bound
\cite{marton1979broadcast,gelfand1980capacity,liang2007broadcast})
and distributed lossy source coding (Berger-Tung inner bound \cite{berger1978multiterminal,tung1978multiterminal}).
Achievability results are proved using the coding theorem by Lee and
Chung \cite{lee2015unified,lee2018unified}, together with a novel
simplification algorithm to reduce the number of auxiliary random
variables via existential information inequalities.

An open-source Python implementation of the framework described in
this paper, called Python Symbolic Information Theoretic Inequality
Prover (PSITIP), is available online\footnote{Source code is available at \href{https://github.com/cheuktingli/psitip}{https://github.com/cheuktingli/psitip}}.
PSITIP is capable of proving 57.1\% (32 out of 56) of the theorems
in Chapters 1-14 of \cite{elgamal2011network} (if we exclude 16 theorems
on Gaussian settings that are out of the scope of our framework, then
PSITIP can prove 80\% of the theorems on discrete settings).\footnote{For the list of theorems and proofs generated by the program, visit
\href{https://github.com/cheuktingli/psitip}{https://github.com/cheuktingli/psitip}} 

This paper is organized as follows. In Section \ref{sec:lp}, we review
the linear programming method for proving information inequalities
in \cite{yeung1996itip,yeung1997framework}. In Section \ref{sec:ei_ineq},
we introduce the notion of existential information inequalities (EIIs).
In Section \ref{sec:auxsearch}, we describe the auxiliary searching
algorithm, a method for proving EIIs. In Section \ref{sec:neii},
we describe some inference rules for EIIs. In Section \ref{sec:nontrivial},
we study some EIIs that cannot be proved by the algorithm (called
nontrivial EIIs). In Section \ref{sec:auxsearch_premise}, we explain
how to incorporate these nontrivial EIIs into the algorithm to allow
it to prove more nontrivial EIIs. In Section \ref{sec:eip}, we introduce
the notion of existential information predicates (EIPs), which is
useful for representing rate regions in network information theory,
and describe how we can compare and simplify EIPs. In Section \ref{sec:dbc},
we demonstrate the algorithm by showing how it can automatically derive
the capacity region of the degraded broadcast channel \cite{bergmans1973random,gallager1974capacity}.
 In Section \ref{sec:list}, we give a list of results that can be
proved by the algorithm.

\medskip{}

\subsection{Related Works}

The inference problem of conditional independence \cite{dawid1979conditional,spohn1980stochastic,mouchart1984note}
is to decide whether a statement on the conditional independence among
some random variables follows from a list of other such statements.
Pearl and Paz \cite{pearl1987graphoids} introduced an axiomatic system,
called the semi-graphoid axioms, which is useful for characterizing
conditional independence structures among random variables. It was
shown by Studen\'y \cite{studeny1989multiinformation} that the semi-graphoid
axioms are incomplete (i.e., unable to prove all valid conditional
independence inferences). Algorithms for conditional independence
inference has been studied, for example, in \cite{geiger1993logical,bouckaert2007racing,niepert2012logical}.
It was shown recently in \cite{li2022ncundecidability} that the inference
problem of conditional independence is undecidable. Also refer to
\cite{kuhne2022entropic} for another proof posted slightly later.

The problem of characterizing linear inequalities among entropy and
mutual information of random variables (called \emph{linear information
inequalities}) was studied by Yeung \cite{yeung1997framework} and
Zhang and Yeung \cite{zhang1997non,zhang1998characterization}. The
Information Theoretic Inequality Prover (ITIP) by Yeung and Yan \cite{yeung1996itip}
is a program that is capable of performing inference on linear information
inequalities (i.e., deducing whether an inequality follows from a
list of other inequalities) via linear programming. For other programs
based on this linear programming method, see \cite{pulikkoonattuxitip,gattegno2016fourier,tian2019user,ho2020proving}.
Refer to \cite{ling2019scalable,ho2020proving} for recent advances
on the linear programming algorithm, and \cite{yeung2021machine}
for an overview. A symbolic approach to proving Shannon-type linear
inequalities was studied in \cite{guo2022proving}. Note that the
inference problem of linear information inequality is a generalization
of the inference problem of conditional independence, since conditional
independence can be expressed using conditional mutual information. 

While ITIP is strictly more powerful than the semi-graphoid axioms
(it can also prove the example given in \cite{studeny1989multiinformation}),
it is still incomplete. It is only capable of proving Shannon-type
inequalities, whereas non-Shannon-type inequalities were discovered
in \cite{zhang1997non,zhang1998characterization,makarychev2002new,dougherty2006six,matus2007infinitely,xu2008projection,dougherty2011non}.
Proofs of non-Shannon-type inequalities often invoke the copy lemma
\cite{zhang1998characterization,dougherty2011non}, and specialized
algorithms employing the copy lemma has been used in \cite{xu2008projection,dougherty2011non}
to discover non-Shannon-type inequalities, and in \cite{farras2018improving,gurpinar2019use}
for other purposes.

Programs for performing inference on information inequalities have
found success in proving results on several settings in information
theory (e.g. network coding \cite{yeung2008information,chan2008dualities,yan2012implicit,apte2014algorithms,apte2016explicit}
and secret sharing \cite{metcalf2011improved,farras2018improving,gurpinar2019use}).
For information theory in general, ITIP is a convenient tool for manipulating
expressions involving entropy and mutual information terms. While
it is useful for intermediate steps, it is often incapable of performing
the whole proof. For example, in network information theory \cite{elgamal2011network},
capacity regions are often stated in terms of auxiliary random variables,
and the existence of auxiliary random variables cannot be stated in
terms of information inequalities. The role of ITIP (and related programs)
in proving these results is akin to a calculator for simple computations
(e.g. for Fourier-Motzkin elimination in \cite{gattegno2016fourier}),
but not a complete package.

While the algorithm in this paper is capable of proving a wide range
of problems in information theory, for many of these classes of problems,
it is fundamentally impossible to have a algorithm that can solve
all problems within the class. For example, it was shown recently
in \cite{li2022ncundecidability} that the problem of conditional
information inequalities is undecidable, i.e., cannot be solved by
an algorithm. Also see \cite{li2021undecidabilityitw,li2021first}
for earlier partial undecidability results. The decidability of unconditional
information inequalities is still open \cite{gomez2014network,gomez2017defining,khamis2020decision,yeung2021machine}. 

\medskip{}

\subsection*{Notations}

The set of nonnegative real numbers is denoted as $\mathbb{R}_{\ge0}$.
The set of positive integers is denoted as $\mathbb{N}$. Given propositions
$P,Q$, the logical conjunction (i.e., AND) is denoted as $P\wedge Q$,
and the material implication is denoted as $P\to Q:=(\mathrm{NOT}\,P)\,\mathrm{OR}\,Q$.
We write $[a..b]:=\{x\in\mathbb{Z}:\,a\le x\le b\}$, $[n]:=[1..n]$.
Given vectors $\mathbf{v},\mathbf{w}\in\mathbb{R}^{k}$, $\mathbf{v}\succeq\mathbf{w}$
means $\mathbf{v}_{i}\ge\mathbf{w}_{i}$ for $i\in[k]$. We use the
same notation for entrywise comparison for matrices.   We write
$X_{a}^{b}:=(X_{a},X_{a+1},\ldots,X_{b})$, $X^{n}:=X_{1}^{n}$. For
finite set $\mathcal{S}\subseteq\mathbb{N}$, write $X_{\mathcal{S}}:=(X_{a_{1}},\ldots,X_{a_{k}})$,
where $a_{1},\ldots,a_{k}$ are the elements of $\mathcal{S}$ in
ascending order. For a statement $s$, its indicator function is written
as $\mathbf{1}\{s\}$ (which is $1$ if $s$ holds, $0$ otherwise). 

We usually use $\mathbf{A},\mathbf{B},\mathbf{C},\mathbf{D}$ for
matrices, $\mathbf{a},\mathbf{b},\mathbf{c},\mathbf{d}$ for column
vectors, $X,Y,Z,U,V,W$ for random variables, $R$ for real (non-random)
variables, and $\mathcal{S},\mathcal{T}$ for sets. We write $\mathbf{1}^{n\times k}$
for the $n\times k$ matrix with ones, and $\mathbf{0}^{n\times k}$
for the $n\times k$ matrix with zeros. Write $\mathbf{I}_{n}$ for
the $n\times n$ identity matrix. For a more compact notation, we
sometimes write $[\mathbf{A};\mathbf{B}]=\left[\begin{array}{c}
\mathbf{A}\\
\mathbf{B}
\end{array}\right]$ for the vertical stacking of two matrices.

\medskip{}

\section{Linear Program\label{sec:lp}}

In this section, we review the linear programming approach in \cite{yeung1996itip,yeung1997framework}.
For a sequence of random variables (or random sequence in short) $X^{n}=(X_{1},\ldots,X_{n})$,
its entropic vector \cite{zhang1997non} is defined as $\mathbf{h}(X^{n})=\mathbf{h}\in\mathbb{R}^{2^{n}-1}$,
where the entries of $\mathbf{h}$ are indexed by nonempty subsets
of $[n]$ (there are $2^{n}-1$ such nonempty subsets\footnote{\label{fn:binary}The index $i\in\{1,\ldots,2^{n}-1\}$ in $\mathbf{h}\in\mathbb{R}^{2^{n}-1}$
corresponds to the subset $\mathcal{S}\subseteq[n]$ that is the binary
representation of $i$, i.e., $i=\sum_{k\in\mathcal{S}}2^{k-1}$.}), and $\mathbf{h}_{\mathcal{S}}:=H(X_{\mathcal{S}})$ (where $\mathcal{S}\subseteq[n]$)
is the joint entropy of $\{X_{i}\}_{i\in\mathcal{S}}$. The entropic
region \cite{zhang1997non} is defined as the region of entropic vectors
\[
\Gamma_{n}^{*}:=\bigcup_{p_{X^{n}}}\{\mathbf{h}(X^{n})\}
\]
over all discrete joint distributions $p_{X^{n}}$. The entropic region
is hard to characterize (the problem of characterizing $\Gamma_{n}^{*}$
for $n\ge4$ is open). Therefore, for the purpose of automated verification
of information-theoretic inequalities, we often focus on Shannon-type
inequalities in the form $I(X;Y|Z)\ge0$, or equivalently $H(X,Z)+H(Y,Z)-H(X,Y,Z)-H(Z)\ge0$.
The set of vectors satisfying Shannon-type inequalities is given by
\cite{zhang1998characterization}
\begin{align*}
\Gamma_{n} & :=\Big\{\mathbf{k}\in\mathbb{R}^{2^{n}-1}:\,\forall\mathcal{S}\subseteq\mathcal{T}:\,\mathbf{k}_{\mathcal{S}}\le\mathbf{k}_{\mathcal{T}}\\
 & \;\;\;\;\;\;\wedge\;\forall\mathcal{S},\mathcal{T}:\,\mathbf{k}_{\mathcal{S}}+\mathbf{k}_{\mathcal{T}}\ge\mathbf{k}_{\mathcal{S}\cup\mathcal{T}}+\mathbf{k}_{\mathcal{S}\cap\mathcal{T}}\Big\},
\end{align*}
where we let $\mathbf{k}_{\emptyset}=0$. Note that $\Gamma_{n}$
is a convex polyhedral cone. The number of constraints in $\Gamma_{n}$
can be reduced by considering only the elemental inequalities \cite{yeung1997framework}.
While we have $\Gamma_{n}^{*}\subseteq\Gamma_{n}$, the inclusion
is strict for $n\ge3$ \cite{zhang1997non,zhang1998characterization}.

The linear program in ITIP \cite{yeung1996itip,yeung1997framework}
attempts to prove results in the form ``for any $X^{n}$ such that
$\mathbf{A}\mathbf{h}(X^{n})\succeq0$ holds, $\mathbf{B}\mathbf{h}(X^{n})\succeq0$
must hold'', i.e.,
\begin{equation}
\forall p_{X^{n}}:\,\left(\mathbf{A}\mathbf{h}(X^{n})\succeq0\,\to\,\mathbf{B}\mathbf{h}(X^{n})\succeq0\right)\label{eq:h_impl}
\end{equation}
for $\mathbf{A}\in\mathbb{R}^{m_{\mathbf{A}}\times(2^{n}-1)}$, $\mathbf{B}\in\mathbb{R}^{m_{\mathbf{B}}\times(2^{n}-1)}$.
Note that inequalities between (linear combinations of) entropy and
mutual information terms involving $X_{1},\ldots,X_{n}$ can always
be expressed in the form $\mathbf{A}\mathbf{h}(X^{n})\succeq0$, since
(conditional) mutual information can be written as a linear combination
of entropy terms. We call \eqref{eq:h_impl} a \emph{conditional information
inequality} (CII). If $\mathbf{A}$ is an empty matrix (i.e., $m_{\mathbf{A}}=0$),
then we also call \eqref{eq:h_impl} an \emph{unconditional information
inequality} (UII). ITIP tries to prove \eqref{eq:h_impl} by solving
the linear program
\begin{equation}
\mathrm{minimize}\;\;\mathbf{b}^{T}\mathbf{k}\;\;\mathrm{subject}\,\mathrm{to}\;\;\mathbf{k}\in\Gamma_{n},\,\mathbf{A}\mathbf{k}\succeq0\label{eq:cii_lp}
\end{equation}
for each row $\mathbf{b}^{T}$ of $\mathbf{B}$, and check whether
the optimal values are non-negative for every row. If this holds,
then the implication in \eqref{eq:h_impl} is true. A limitation of
ITIP is that it can only verify Shannon-type inequalities (since it
uses $\Gamma_{n}$ rather than $\Gamma_{n}^{*}$), i.e., it can only
find the truth value of
\begin{equation}
\forall\mathbf{k}\in\Gamma_{n}:\,\left(\mathbf{A}\mathbf{k}\succeq0\,\to\,\mathbf{B}\mathbf{k}\succeq0\right),\label{eq:h2_impl}
\end{equation}
which is a sufficient, but not necessary condition for \eqref{eq:h_impl}
to hold (i.e., it cannot prove that \eqref{eq:h_impl} does not hold).
Examples of $\mathbf{A},\mathbf{B}$ where \eqref{eq:h_impl} holds
but \eqref{eq:h2_impl} does not are called \emph{non-Shannon-type}
CIIs. A non-Shannon-type CII is given in \cite{zhang1997non}, whereas
non-Shannon-type UIIs are given in \cite{zhang1998characterization,makarychev2002new,dougherty2006six,matus2007infinitely,xu2008projection,dougherty2011non}.
\begin{rem}
Conventionally, in an information inequality (UII or CII), there is
only one inequality in the consequence in \eqref{eq:h_impl}, i.e.,
$\mathbf{B}$ has one row. It is indeed sufficient to consider the
case where $\mathbf{B}$ has one row, since the case for multiple
rows is equivalent to the conjunction of the CIIs for all rows (similar
to \eqref{eq:cii_lp}). However, in this paper, we consider the general
case where $\mathbf{B}$ can have multiple rows, which is crucial
to existential information inequalities defined in the next section.
\end{rem}
\medskip{}

\section{Existential Information Inequalities\label{sec:ei_ineq}}

The random sequence $X^{n}$ in \eqref{eq:h_impl} is quantified with
universal quantification (i.e, ``$\forall$''). In this section,
we will extend \eqref{eq:h_impl} to incorporate existential quantification.
\emph{Existential information inequalities} (EII) are statements in
the form:
\begin{equation}
\forall p_{X^{n}}\!:\left(\mathbf{A}\mathbf{h}(X^{n})\succeq0\,\to\,\exists p_{U^{l}|X^{n}}\!:\mathbf{B}\mathbf{h}(X^{n},U^{l})\succeq0\right),\label{eq:aux_impl}
\end{equation}
where $U^{l}$ are called the auxiliary random variables (or ``auxiliary''
in short), and $p_{U^{l}|X^{n}}$ is their conditional distribution
given $X^{n}$. Note that $\mathbf{h}(X^{n},U^{l})\in\mathbb{R}^{2^{n+l}-1}$
is the entropic vector of the sequence of random variables $(X_{1},\ldots,X_{n},U_{1},\ldots,U_{l})$.
Also note that $\mathbf{A}\in\mathbb{R}^{m_{\mathbf{A}}\times(2^{n}-1)}$,
$\mathbf{B}\in\mathbb{R}^{m_{\mathbf{B}}\times(2^{n+l}-1)}$. For
brevity, we will write \eqref{eq:aux_impl} as
\begin{equation}
\forall X^{n}:\,\left(\mathbf{A}\mathbf{h}(X^{n})\succeq0\,\to\,\exists U^{l}:\,\mathbf{B}\mathbf{h}(X^{n},U^{l})\succeq0\right),\label{eq:aux_impl2}
\end{equation}
which reads as ``for all $X^{n}$ satisfying $\mathbf{A}\mathbf{h}(X^{n})\succeq0$,
there exists $U^{l}$ satisfying $\mathbf{B}\mathbf{h}(X^{n},U^{l})\succeq0$''.
When a new random variable is declared (e.g. ``$\forall X^{n}$''
and ``$\exists U^{l}$''), it is assumed to be dependent on all
previously declared random variables (e.g. they are shorthands for
``$\forall p_{X^{n}}$'' and ``$\exists p_{U^{l}|X^{n}}$'' respectively).
The sets of UIIs, CIIs and EIIs satisfy the following inclusion: UIIs
$\subseteq$ CIIs $\subseteq$ EIIs (EII is the most general since
it reduces to CII when $l=0$). As a corollary of the undecidability
of CII shown recently in \cite{li2022ncundecidability}, EII (when
$\mathbf{A}$, $\mathbf{B}$ have rational entries) is undecidable
as well. Nevertheless, we will discuss several algorithms that can
verify EIIs that arise from a wide range of problems.

There are several situations where EIIs arise. For example:
\begin{enumerate}
\item In network information theory, capacity regions are often stated in
terms of auxiliary random variables (see various examples in \cite{elgamal2011network}).
We will show that a capacity region can often be expressed as an existential
information predicate (defined in Section \ref{sec:eip}). Both converse
and achievability proofs can be performed under the framework of EII:
\begin{enumerate}
\item (Converse). The most common technique for proving outer bounds and
converse results pioneered by Gallager \cite{gallager1974capacity}
involves writing the $n$-letter operational region of the coding
setting, and identifying the auxiliary random variables (among the
variables appearing in the operational region) so that the proposed
outer bound is satisfied. The problem of finding auxiliaries satisfying
some conditions can often be represented as an EII. Refer to Remark
\ref{rem:nit} for a discussion, Section \ref{sec:dbc} for a demonstration
on the degraded broadcast channel, and Section \ref{sec:list} for
a non-exhaustive list of results provable by the framework. \medskip{}
\item (Achievability). Several general achievability results combining techniques
such as superposition coding, simultaneous nonunique decoding and
binning have been studied (e.g. \cite{rini2013general,minero2015unified,lee2015unified}).
A downside of these results is that the inner bounds they give are
often complicated (contain more auxiliaries than needed), and need
to be simplified manually to obtain the final inner bound. Moreover,
these results include some parameters that are often chosen manually
(e.g. which messages should a decoder decode uniquely and nonuniquely),
and we often have to compare inner bounds obtained using different
parameters in order to select the largest one. The framework in this
paper provides a systematic method to simplify and compare rate regions.
The PSITIP implementation allows automated discovery of inner bounds,
using only the graphical representation of the network as input, via
the inner bound in \cite{lee2015unified,lee2018unified} together
with the simplification procedures in Section \ref{sec:eip}.\medskip{}
\end{enumerate}
\item Auxiliary random variables are used in the definitions of some information-theoretic
quantities such as Wyner's common information \cite{wyner1975common}
(the $U$ is an auxiliary random variable)
\[
J(X;Y):=\inf_{p_{U|X,Y}:I(X;Y|U)=0}I(U;X,Y),
\]
 and G{\'a}cs-K{\"o}rner common information \cite{gacs1973common}
\[
K(X;Y):=\sup_{p_{U|X,Y}:H(U|X)=H(U|Y)=0}H(U).
\]
Other examples include common entropy \cite{kumar2014exact,li2017distributed},
necessary conditional entropy \cite{cuff2010coordination}, information
bottleneck \cite{tishby2000information}, privacy funnel \cite{makhdoumi2014information},
excess functional information \cite{sfrl_trans}, and other quantities
studied in \cite{li2017extended}. Properties of these quantities
can often be stated as EIIs (see Example \ref{exa:wyner}). In fact,
for information quantities in the form 
\begin{equation}
F(X^{n}):=\inf_{U^{l}:\mathbf{B}\mathbf{h}(X^{n},U^{l})\succeq0}\mathbf{b}^{T}\mathbf{h}(X^{n},U^{l}),\label{eq:infoquantity}
\end{equation}
where $\mathbf{B}\in\mathbb{R}^{m_{\mathbf{B}}\times(2^{n+l}-1)}$
and $\mathbf{b}\in\mathbb{R}^{2^{n+l}-1}$ (which includes all aforementioned
information quantities; $\sup$ can be expressed as $\inf$ by flipping
the sign), any linear inequality among such information quantities
can be expressed as an EII. This can be shown by observing
\begin{align*}
 & \forall X^{n}:\Big(\mathbf{A}\mathbf{h}(X^{n})\succeq0\,\to\\
 & \;\;\inf_{U^{l}:\mathbf{B}\mathbf{h}(X^{n},U^{l})\succeq0}\!\!\mathbf{b}^{T}\mathbf{h}(X^{n},U^{l})\ge\!\!\inf_{V^{k}:\mathbf{C}\mathbf{h}(X^{n},V^{k})\succeq0}\!\!\mathbf{c}^{T}\mathbf{h}(X^{n},V^{k})\Big)\\
 & \Leftrightarrow\,\forall X^{n}U^{l}:\Big(\mathbf{A}\mathbf{h}(X^{n})\succeq0\,\wedge\,\mathbf{B}\mathbf{h}(X^{n},U^{l})\succeq0\,\to\\
 & \;\;\exists V^{k}:\mathbf{C}\mathbf{h}(X^{n},V^{k})\succeq0\,\wedge\,\mathbf{b}^{T}\mathbf{h}(X^{n},U^{l})\ge\mathbf{c}^{T}\mathbf{h}(X^{n},V^{k})\Big).
\end{align*}
Note that linear inequalities among more than two information quantities
(in the form \eqref{eq:infoquantity}) can also be stated in this
form, since we can combine all positive terms into one, and all negative
terms into one, resulting in at most two information quantities. Therefore,
by employing EII, we can handle linear inequalities not only among
entropy and mutual information terms, but also on more general information
quantities.\medskip{}
\item In the study of the entropic region $\Gamma_{n}^{*}$, outer bounds
(e.g. $\Gamma_{n}$ and various non-Shannon-type inequalities) can
be stated as UIIs. Nevertheless, inner bounds of $\Gamma_{n}^{*}$
(e.g. \cite{chan2002relation,hassibi2007construction,walsh2011recursive})
cannot be stated as UIIs, though they can be stated as EIIs. For example,
the fact that $\Gamma_{2}$ is an inner bound of $\Gamma_{2}^{*}$
(actually $\Gamma_{2}=\Gamma_{2}^{*}$ as shown in \cite{zhang1997non})
can be expressed as
\begin{align*}
 & \forall R_{1},R_{2},R_{3}\ge0:\big(R_{1}\le R_{3}\,\wedge\,R_{2}\le R_{3}\\
 & \wedge\,R_{3}\le R_{1}+R_{2}\,\to\exists U^{2}:\\
 & H(U_{1})=R_{1}\,\wedge\,H(U_{2})=R_{2}\,\wedge\,H(U_{1},U_{2})=R_{3}\big).
\end{align*}
Note that $R_{1},R_{2},R_{3}$ are real-valued non-random variables.
Refer to Remark \ref{rem:real} for how to represent them in an EII.
Combine this with the fact that the outer bound $\Gamma_{2}^{*}$
can be expressed as a UII, the statement $\Gamma_{2}=\Gamma_{2}^{*}$
can also be expressed as an EII.\medskip{}
\item The copy lemma \cite{zhang1998characterization,dougherty2011non},
a useful tool for proving non-Shannon-type inequalities, can be stated
as an EII. Refer to Section \ref{subsec:copylem} for details.\medskip{}
\end{enumerate}
The following example demonstrates the usage of EIIs.
\begin{example}
\label{exa:wyner}Consider the tensorization property of Wyner's common
information, i.e., $(X_{1},Y_{1})\perp\!\!\!\perp(X_{2},Y_{2})$ implies
$J(X_{1},X_{2};Y_{1},Y_{2})=J(X_{1};Y_{1})+J(X_{2};Y_{2})$. We can
prove inequalities in both directions:
\begin{enumerate}
\item $J(X_{1},X_{2};Y_{1},Y_{2})\ge J(X_{1};Y_{1})+J(X_{2};Y_{2})$ is
equivalent to  the EII
\begin{align*}
 & \forall X^{2},Y^{2},V\!:\!\bigg(\!I(X_{1},Y_{1};X_{2},Y_{2})=I(X_{1},X_{2};Y_{1},Y_{2}|V)=0\\
 & \rightarrow\exists U^{2}:\,I(X_{1};Y_{1}|U_{1})=I(X_{2};Y_{2}|U_{2})=0\,\\
 & \;\;\;\;\;\;\wedge\;I(V;X^{2},Y^{2})\ge I(U_{1};X_{1},Y_{1})+I(U_{2};X_{2},Y_{2})\bigg).
\end{align*}
This can be proved by identifying $U_{1}=U_{2}=V$.
\item $J(X_{1},X_{2};Y_{1},Y_{2})\le J(X_{1};Y_{1})+J(X_{2};Y_{2})$ is
equivalent to the EII
\begin{align*}
 & \forall X^{2},Y^{2},U^{2}:\bigg(I(X_{1},Y_{1},U_{1};X_{2},Y_{2},U_{2})\\
 & \;\;=I(X_{1};Y_{1}|U_{1})=I(X_{2};Y_{2}|U_{2})=0\,\rightarrow\\
 & \exists V:\,I(X^{2};Y^{2}|V)=0\;\\
 & \;\wedge\;I(V;X^{2},Y^{2})\le I(U_{1};X_{1},Y_{1})+I(U_{2};X_{2},Y_{2})\bigg).
\end{align*}
This can be proved by identifying $V=(U_{1},U_{2})$ (the joint random
variable). We can assume $I(X_{1},Y_{1},U_{1};X_{2},Y_{2},U_{2})=0$
since $J(X_{1};Y_{1})$ and $J(X_{2};Y_{2})$ can be optimized separately.
If we do not make this assumption (i.e., we only have $I(X_{1},Y_{1};X_{2},Y_{2})=0$),
then the EII can still be proved using the conditional independence
rule in Section \ref{subsec:copylem}.
\end{enumerate}
\end{example}
\medskip{}

\begin{rem}
\label{rem:real}In \eqref{eq:aux_impl2}, the variables $X_{i},U_{i}$
are random variables. We may also want to introduce real-valued non-random
variables (real variables in short), e.g. rates in a coding setting.
For example, we have $\forall R\ge0:\,\exists U:\,H(U)=R/2$ (i.e.,
for any real number $R\ge0$, there exists random variable $U$ with
$H(U)=R/2$). To represent this statement in the form \eqref{eq:aux_impl2},
we replace each nonnegative real variable $R$ by $H(X)$ for a new
random variable $X$, i.e., it becomes $\forall X:\,\exists U:\,H(U)=H(X)/2$.
In case $R$ is not constrained to be nonnegative, replace $R$ by
$H(X_{1})-H(X_{2})$ for two new random variables $X_{1},X_{2}$.
This way we can transform a statement involving real variables into
an equivalent EII. Nevertheless, since the dimension of the linear
program is exponential in the number of random variables, this method
is inefficient in practice. In the PSITIP implementation, real variables
are represented separately.
\end{rem}
\medskip{}

\begin{rem}
\label{rem:nit} We discuss a method to state a converse result in
network information theory as an EII, which is basically Gallager's
approach \cite{gallager1974capacity}. For each sequence of random
variables $X^{n}$ in the $n$-letter operational setting, define
a group of three random variables $X_{Q},X^{Q-1},X_{Q+1}^{n}$, representing
the present, past and future respectively, where $Q\sim\mathrm{Unif}[n]$
is the time sharing random variable. The causal relations between
the random variables and the decoding requirements are included in
the EII. Csisz{\'a}r sum identity \cite{korner1977images,csiszar1978broadcast}
applied on each pair of those groups is added to the condition in
the EII. This way, the number of variables in the EII is only 3 times
the number of variables in the 1-letter operational setting, and does
not grow with $n$. This strategy suffices for most of the converse
and outer bound proofs in \cite{elgamal2011network} (some results
that have been verified by PSITIP are listed in Section \ref{sec:list}).

We will not only discuss an algorithm that proves a given outer bound,
but also an algorithm that automatically produces an outer bound given
a description of the coding setting. The aforementioned method, which
considers the present, past and future versions of each random sequence
(where the past and future are regarded as auxiliaries), creates too
many auxiliaries, making the resultant outer bound complicated and
hard to interpret. To produce a satisfactory outer bound, we describe
algorithms that reduce the number of auxiliaries. The algorithms basically
automate the process of identifying new auxiliaries as combinations
of existing auxiliaries (e.g. past and future variables), which was
performed manually in conventional converse proofs. They are discussed
in Section \ref{sec:eip}. 
\end{rem}
\medskip{}

\begin{rem}
The PSITIP implementation is capable of handling arbitrary first-order
statement on entropy, i.e., composition of linear inequalities on
entropy and mutual information, existential and universal quantification
of random variables, AND, OR and NOT operators (it can handle statements
with an arbitrary sequence of ``$\forall$'' and ``$\exists$'',
not only that in the form of \eqref{eq:aux_impl}), though we will
not discuss this here for simplicity.
\end{rem}
\medskip{}

\section{Substitution Operation}

As demonstrated in Example \ref{exa:wyner}, to prove an EII, we can
identify the auxiliary random variables using the non-auxiliary random
variables. To prove \eqref{eq:aux_impl}, a sufficient condition is
to identify $\mathcal{S}_{1},\mathcal{S}_{2},\ldots,\mathcal{S}_{l}\subseteq[n]$
such that when we substitute $U_{1}=X_{\mathcal{S}_{1}},\ldots,U_{l}=X_{\mathcal{S}_{l}}$,
we have 
\begin{equation}
\mathbf{A}\mathbf{h}(X^{n})\succeq0\,\rightarrow\,\mathbf{B}\mathbf{h}(X^{n},U^{l})\succeq0,\label{eq:sub_impl_ex}
\end{equation}
which can be checked using linear programming after we fix $\mathcal{S}_{1},\ldots,\mathcal{S}_{l}$.
\footnote{Note that even though $U_{i}$ can be joint random variables, it is
considered as one random variable in $\mathbf{h}(X^{n},U^{l})$, and
we still have $\mathbf{h}(X^{n},U^{l})\in2^{n+l}-1$. Nevertheless,
since $X^{n}$ is the only randomness ($U_{i}$ are functions of $X^{n}$),
each entry of $\mathbf{h}(X^{n},U^{l})$ is also an entry of $\mathbf{h}(X^{n})$.} 

A way of substituting the auxiliaries $U^{l}$ by the variables $X^{n}$
can be specified by a sequence of sets $\mathcal{S}_{1},\mathcal{S}_{2},\ldots,\mathcal{S}_{l}\subseteq[n]$,
where we substitute $U_{i}=X_{\mathcal{S}_{i}}$. We call $(\mathcal{S}_{1},\mathcal{S}_{2},\ldots,\mathcal{S}_{l})$
a \emph{substitution combination}. We introduce a more compact way
to represent a substitution combination using a matrix $\mathbf{S}\in\mathbb{R}_{\ge0}^{l\times n}$
of nonnegative entries, where 
\[
\mathbf{S}_{i,j}=\begin{cases}
1 & \mathrm{if}\;j\in\mathcal{S}_{i}\\
0 & \mathrm{if}\;j\neq\mathcal{S}_{i}.
\end{cases}
\]
We call $\mathbf{S}$ the \emph{substitution matrix} corresponding
to $(\mathcal{S}_{1},\mathcal{S}_{2},\ldots,\mathcal{S}_{l})$. Given
a random sequence $X^{n}$ and a matrix $\mathbf{S}\in\mathbb{R}_{\ge0}^{l\times n}$,
we write $\mathbf{S}\circ X^{n}$ for the random sequence with length
$l$, where the $i$-th entry is
\[
(\mathbf{S}\circ X^{n})_{i}:=X_{\{j\in[n]:\,\mathbf{S}_{i,j}\neq0\}}.
\]
We call ``$\circ$'' the \emph{substitution operation}. If $\mathbf{S}$
is the substitution matrix corresponding to $(\mathcal{S}_{1},\mathcal{S}_{2},\ldots,\mathcal{S}_{l})$,
then $\mathbf{S}\circ X^{n}$ is precisely the sequence $U^{l}$ where
$U_{i}=X_{\mathcal{S}_{i}}$. Note that the substitution operation
does not require the matrix to have $\{0,1\}$ entries (it only requires
nonnegative entries), and it only depends on the positions of the
nonzero entries in $\mathbf{S}$, not their precise values. We can
show that the substitution operation satisfies the following associativity
law.
\begin{prop}
[Associativity of substitution operation]For any random sequence
$X^{n}$, and matrices $\mathbf{S}\in\mathbb{R}_{\ge0}^{l\times n}$,
$\mathbf{T}\in\mathbb{R}_{\ge0}^{m\times l}$, we have 
\[
(\mathbf{T}\mathbf{S})\circ X^{n}\stackrel{\iota}{=}\mathbf{T}\circ(\mathbf{S}\circ X^{n}),
\]
where $Y^{m}\stackrel{\iota}{=}Z^{m}$ means $H(Y_{i}|Z_{i})=H(Z_{i}|Y_{i})=0$
for $i\in[m]$, i.e., the two random sequences are informationally
equivalent. 
\end{prop}
\begin{IEEEproof}
This can be shown by 
\begin{align*}
((\mathbf{T}\mathbf{S})\circ X^{n})_{i} & =X_{\{j\in[n]:\,(\mathbf{T}\mathbf{S})_{i,j}\neq0\}}\\
 & =X_{\{j\in[n]:\,\sum_{k=1}^{l}\mathbf{T}_{i,k}\mathbf{S}_{k,j}\neq0\}}\\
 & =X_{\{j\in[n]:\,\exists k\in[l]:\,\mathbf{T}_{i,k}\mathbf{S}_{k,j}\neq0\}}\\
 & \stackrel{\iota}{=}U_{\{k\in[l]:\,\mathbf{T}_{i,k}\neq0\}},
\end{align*}
where $U_{k}:=X_{\{j\in[n]:\,\mathbf{S}_{k,j}\neq0\}}$. 
\end{IEEEproof}
\medskip{}

We now describe how the substitution operation affects the entropic
vector. Given a matrix $\mathbf{S}\in\mathbb{R}_{\ge0}^{l\times n}$,
our goal is to find a matrix $\mathrm{sub}(\mathbf{S})\in\{0,1\}^{(2^{l}-1)\times(2^{n}-1)}$,
called the \emph{entropic substitution matrix} of $\mathbf{S}$, such
that for any random sequence $X^{n}$, we have
\[
\mathbf{h}(\mathbf{S}\circ X^{n})=\mathrm{sub}(\mathbf{S})\mathbf{h}(X^{n}).
\]
This is possible since each entry of $\mathbf{h}(\mathbf{S}\circ X^{n})$
is also an entry of $\mathbf{h}(X^{n})$. More precisely, the $\mathcal{T}$-th
entry of $\mathbf{h}(\mathbf{S}\circ X^{n})$ (where $\mathcal{T}\subseteq[l]$)
is 
\begin{align*}
H\big((\mathbf{S}\circ X^{n})_{\mathcal{T}}\big) & =H\big(X_{\bigcup_{i\in\mathcal{T}}\{j\in[n]:\,\mathbf{S}_{i,j}\neq0\}}\big)\\
 & =H\big(X_{\{j\in[n]:\,\sum_{i\in\mathcal{T}}\mathbf{S}_{i,j}\neq0\}}\big).
\end{align*}
Hence, we can take 
\begin{equation}
(\mathrm{sub}(\mathbf{S}))_{\mathcal{T},\mathcal{U}}:=\begin{cases}
1 & \mathrm{if}\;\mathcal{U}=\big\{ j\in[n]:\,\sum_{i\in\mathcal{T}}\mathbf{S}_{i,j}\neq0\big\}\\
0 & \mathrm{if}\;\mathcal{U}\neq\big\{ j\in[n]:\,\sum_{i\in\mathcal{T}}\mathbf{S}_{i,j}\neq0\big\}
\end{cases}\label{eq:def_comb}
\end{equation}
for nonempty $\mathcal{T}\subseteq[l]$, $\mathcal{U}\subseteq[n]$.
Note that $(\mathrm{sub}(\mathbf{S}))_{\mathcal{T},\mathcal{U}}$
denotes the entry of $\mathrm{sub}(\mathbf{S})$ in the $\mathcal{T}$-th
row and the $\mathcal{U}$-th column (like entropic vectors, the rows
and columns of this matrix are also indexed by nonempty subsets). 

By the associativity of the substitution operation, the function $\mathrm{sub}$
is multiplicative.
\begin{prop}
[Multiplicativity of entropic substitution matrix]\label{prop:sub_mul}For
any matrices $\mathbf{S}\in\mathbb{R}_{\ge0}^{l\times n}$, $\mathbf{T}\in\mathbb{R}_{\ge0}^{m\times l}$,
we have 
\[
\mathrm{sub}(\mathbf{T})\mathrm{sub}(\mathbf{S})=\mathrm{sub}(\mathbf{T}\mathbf{S}).
\]
\end{prop}
\begin{IEEEproof}
This can be shown by noting that for any $X^{n}$,
\begin{align*}
\mathrm{sub}(\mathbf{T}\mathbf{S})\mathbf{h}(X^{n}) & =\mathbf{h}((\mathbf{T}\mathbf{S})\circ X^{n})\\
 & =\mathbf{h}(\mathbf{T}\circ(\mathbf{S}\circ X^{n}))\\
 & =\mathrm{sub}(\mathbf{T})\mathbf{h}(\mathbf{S}\circ X^{n})\\
 & =\mathrm{sub}(\mathbf{T})\mathrm{sub}(\mathbf{S})\mathbf{h}(X^{n}),
\end{align*}
and that $\mathrm{span}\bigcup_{p_{X^{n}}}\{\mathbf{h}(X^{n})\}=\mathbb{R}^{2^{n}-1}$
by considering $U\sim\mathrm{Bern}(1/2)$, $X_{i}=U$ if $i\in\mathcal{S}$,
$X_{i}=0$ if $i\notin\mathcal{S}$, where each choice of $\mathcal{S}\subseteq[n]$,
$\mathcal{S}\neq\emptyset$ gives a linearly independent $\mathbf{h}(X^{n})$.
\end{IEEEproof}
\medskip{}

For the substitution performed in \eqref{eq:sub_impl_ex}, letting
$\mathbf{S}\in\mathbb{R}_{\ge0}^{l\times n}$ be the substitution
matrix corresponding to $(\mathcal{S}_{1},\mathcal{S}_{2},\ldots,\mathcal{S}_{l})$,
we take $\mathbf{C}=\mathbf{B}\mathrm{sub}(\mathbf{I}_{n};\mathbf{S})$,
where we write
\[
\mathrm{sub}(\mathbf{I}_{n};\mathbf{S})=\mathrm{sub}([\mathbf{I}_{n};\mathbf{S}])=\mathrm{sub}\left(\left[\begin{array}{c}
\mathbf{I}_{n}\\
\mathbf{S}
\end{array}\right]\right)
\]
for a more compact notation. When $U_{i}=X_{\mathcal{S}_{i}}$, we
have $[\mathbf{I}_{n};\mathbf{S}]\circ X^{n}=(X^{n},U^{l})$ (where
$(X^{n},U^{l})$ is a random sequence with length $n+l$) and hence
$\mathbf{B}\mathbf{h}(X^{n},U^{l})=\mathbf{C}\mathbf{h}(X^{n})$.
We can then verify \eqref{eq:sub_impl_ex} by checking $\mathbf{A}\mathbf{h}(X^{n})\succeq0\,\rightarrow\,\mathbf{C}\mathbf{h}(X^{n})\succeq0$
using the linear program in \eqref{eq:cii_lp}. Hence, a method to
verify an EII is to find $\mathbf{S}\in\{0,1\}^{l\times n}$ such
that the linear program succeeds in checking $\mathbf{A}\mathbf{h}(X^{n})\succeq0\,\rightarrow\,\mathbf{C}\mathbf{h}(X^{n})\succeq0$.
The class of EIIs that can be proved this way can be defined formally
as follows.
\begin{defn}
An EII $\forall X^{n}:\,(\mathbf{A}\mathbf{h}(X^{n})\succeq0\,\to\,\exists U^{l}:\,\mathbf{B}\mathbf{h}(X^{n},U^{l})\succeq0)$
is called a \emph{trivial EII} if there exists a substitution matrix
$\mathbf{S}\in\{0,1\}^{l\times n}$ such that the following implication
holds: 
\begin{equation}
\forall\mathbf{k}\in\Gamma_{n}:\,\big(\mathbf{A}\mathbf{k}\succeq0\,\to\,\mathbf{B}\mathrm{sub}(\mathbf{I}_{n};\mathbf{S})\mathbf{k}\succeq0\big),\label{eq:trivial_imp}
\end{equation}
which can be checked using the linear program in \eqref{eq:cii_lp}.
\end{defn}
\medskip{}

Even though trivial EIIs are trivial in a sense, searching for $\mathbf{S}\in\{0,1\}^{l\times n}$
(with $2^{nl}$ different choices) is still a computationally intensive
task. An algorithm for finding $\mathbf{S}\in\{0,1\}^{l\times n}$
using some pruning techniques will be discussed in Section \ref{sec:auxsearch}.

\medskip{}

\section{Verification of EII via Auxiliary Searching\label{sec:auxsearch}}

In this section, we present an algorithm for verifying an EII in the
form \eqref{eq:aux_impl2}, that is capable of verifying any trivial
EII \eqref{eq:trivial_imp}. A simple brute-force algorithm is to
identify $U_{i}=X_{\mathcal{S}_{i}}$, and exhaust all choices of
$\mathcal{S}_{1},\mathcal{S}_{2},\ldots,\mathcal{S}_{l}\subseteq[n]$,
or equivalently exhaust all possible substution matrices $\mathbf{S}\in\{0,1\}^{l\times n}$,
which can be computationally prohibitive. We discuss methods to reduce
the search space. The algorithm is divided into two steps: the sandwich
procedure, and exhaustion with cached linear program optimization.

\medskip{}

\subsection{Sandwich Procedure}

For $\mathbf{b}\in\mathbb{R}^{2^{n}-1}$, we say that $\mathbf{b}^{T}\mathbf{h}(X^{n})$
is \emph{increasing} in $X_{i}$ ($i\in[n]$) if 
\[
\forall X^{n},Y:\left(\mathbf{b}^{T}\mathbf{h}(X^{n})\le\mathbf{b}^{T}\mathbf{h}\big(X^{i-1},(X_{i},Y),X_{i+1}^{n}\big)\right).
\]
The case for decreasing is defined similarly (with ``$\le$'' replaced
with ``$\ge$''). There are several ways of checking whether $\mathbf{b}^{T}\mathbf{h}(X^{n})$
is increasing/decreasing in $X_{i}$. One way is to check the above
condition (which is an UII in the form of \eqref{eq:h_impl}) using
the linear program \eqref{eq:cii_lp}. We call $\mathbf{b}^{T}\mathbf{h}(X^{n})$
\emph{trivially increasing} in $X_{i}$ if it can be verified this
way. In case the expression $\mathbf{b}^{T}\mathbf{h}(X^{n})$ is
given in the form of a linear combination of entropy and mutual information
terms, we can also make use of simple rules like $H(X|Z)$ is increasing
in $X$ and decreasing in $Z$, and $I(X;Y|Z)$ is increasing in $X$
and $Y$ (but neither increasing nor decreasing in $Z$), to deduce
whether the expression is increasing or decreasing.

If $\mathbf{b}^{T}\mathbf{h}(X^{n},U^{l})$ is trivially increasing
in $U_{1}$, then in order for the linear program \eqref{eq:cii_lp}
to be able to verify the CII $\mathbf{A}\mathbf{h}(X^{n})\succeq0\to\mathbf{b}^{T}\mathbf{h}(X^{n},U^{l})\ge0$
for any choice of $\{\mathcal{S}_{i}\}_{i}$ (recall that we try to
identify $U_{i}=X_{\mathcal{S}_{i}}$), it is necessary that \eqref{eq:cii_lp}
can verify the CII when $\mathcal{S}_{1}=[n]$, which gives the largest
$\mathbf{b}^{T}\mathbf{h}(X^{n},U^{l})$. By extension, in order for
\eqref{eq:cii_lp} to be able to verify the CII for any choice of
$\{\mathcal{S}_{i}\}_{i}$ satisfying $1\notin\mathcal{S}_{1}$, it
is necessary that \eqref{eq:cii_lp} can verify the CII when $\mathcal{S}_{1}=[n]\backslash\{1\}$
and $\mathcal{S}_{i}=[n]$ for $i\neq1$. Taking the contrapositive,
if \eqref{eq:cii_lp} fails to verify the CII for this choice of $\{\mathcal{S}_{i}\}_{i}$,
then we can assume $1\in\mathcal{S}_{1}$, giving a lower bound $\{1\}$
on $\mathcal{S}_{1}$ (with respect to set inclusion, i.e., $\{1\}\subseteq\mathcal{S}_{1}$).

If $\mathbf{b}^{T}\mathbf{h}(X^{n},U^{l})$ is trivially increasing
for some $U_{i}$, and trivially decreasing for some other $U_{i}$,
then the most conservative choice will be $\mathcal{S}_{i}=[n]$ for
$U_{i}$ increasing, $\mathcal{S}_{i}=\emptyset$ for $U_{i}$ decreasing.
If we know beforehand that $\underline{\mathcal{S}}_{i}\subseteq\mathcal{S}_{i}\subseteq\overline{\mathcal{S}}_{i}$
for $i\in[l]$, where $\underline{\mathcal{S}}_{i}\subseteq\overline{\mathcal{S}}_{i}\subseteq[n]$
are lower and upper bounds for $\mathcal{S}_{i}$ respectively, then
the most conservative choice will be $\mathcal{S}_{i}=\overline{\mathcal{S}}_{i}$
for $U_{i}$ increasing, $\mathcal{S}_{i}=\underline{\mathcal{S}}_{i}$
for $U_{i}$ decreasing. In terms of the substitution matrix $\mathbf{S}\in\{0,1\}^{l\times n}$,
if we know beforehand that $\underline{\mathbf{S}}\preceq\mathbf{S}\preceq\overline{\mathbf{S}}$,
where $\underline{\mathbf{S}},\overline{\mathbf{S}}\in\{0,1\}^{l\times n}$,
then the most conservative choice for the substitution matrix is
\[
\mathrm{diag}\left(\frac{\mathbf{1}-\mathbf{g}}{2}\right)\underline{\mathbf{S}}+\mathrm{diag}\left(\frac{\mathbf{1}+\mathbf{g}}{2}\right)\overline{\mathbf{S}},
\]
where $\mathbf{g}\in\mathbb{R}^{l}$ with $\mathbf{g}_{i}=1$ if $\mathbf{b}^{T}\mathbf{h}(X^{n},U^{l})$
is trivially increasing for $U_{i}$, and $\mathbf{g}_{i}=-1$ if
trivially decreasing (if it is not trivially increasing or decreasing
for some $i$, we ignore this row $\mathbf{b}^{T}$ in the sandwich
procedure). Therefore, we can repeat this procedure for each row
$\mathbf{b}^{T}$ of $\mathbf{B}$ to refine the lower and upper bounds
on $\mathbf{S}$, reducing the search space for $\mathbf{S}$. The
description of the algorithm for the EII \eqref{eq:aux_impl2} is
given in Algorithm \ref{alg:sandwich}.

\begin{algorithm}[H]
\textbf{$\;\;\;\;$Input:} $\mathbf{A}\in\mathbb{R}^{m_{\mathbf{A}}\times(2^{n}-1)}$,
$\mathbf{B}\in\mathbb{R}^{m_{\mathbf{B}}\times(2^{n+l}-1)}$, initial
values of $\underline{\mathbf{S}},\overline{\mathbf{S}}\in\{0,1\}^{l\times n}$

\textbf{$\;\;\;\;\;\;$} (lower and upper bound of $\mathbf{S}$ respectively,
default values are $\underline{\mathbf{S}}=\mathbf{0}^{l\times n}$,
$\overline{\mathbf{S}}=\mathbf{1}^{l\times n}$)

\textbf{$\;\;\;\;$Output:} final values of $\underline{\mathbf{S}},\overline{\mathbf{S}}\in\{0,1\}^{l\times n}$
or failure

\smallskip{}

\begin{algorithmic}[1]

\Repeat

\For{row $\mathbf{b}^{T}$ of $\mathbf{B}$}

\For{$i\in[l]$}

\State{$\mathbf{g}_{i}\!\leftarrow\!\begin{cases}
1 & \!\!\text{if \ensuremath{\mathbf{b}^{T}}\ensuremath{\mathbf{h}}(\ensuremath{X^{n}},\ensuremath{U^{l}}) triv. increasing in \ensuremath{U_{i}}}\\
-1 & \!\!\text{else if triv. decreasing in \ensuremath{U_{i}}}\\
0 & \!\!\text{otherwise}
\end{cases}$}

\State{\textbf{if} $\mathbf{g}_{i}=0$, \textbf{then} skip row $\mathbf{b}^{T}$
and continue to next row}

\EndFor

\State{$\mathbf{S}\leftarrow\mathrm{diag}\left(\frac{\mathbf{1}-\mathbf{g}}{2}\right)\underline{\mathbf{S}}+\mathrm{diag}\left(\frac{\mathbf{1}+\mathbf{g}}{2}\right)\overline{\mathbf{S}}$}\label{step:stildei}

\State{$\mathbf{c}^{T}\leftarrow\mathbf{b}^{T}\mathrm{sub}(\mathbf{I}_{n};\mathbf{S})$} 

\State{\textbf{if} CII $\forall X^{n}:(\mathbf{A}\mathbf{h}(X^{n})\succeq0\to\mathbf{c}^{T}\mathbf{h}(X^{n})\ge0)$
}

\State{$\;\;\;\;$cannot be verified by \eqref{eq:cii_lp}, \textbf{then}
return failure}

\State{$\;\;\;\;$(note that $\mathbf{c}^{T}\mathbf{h}(X^{n})=\mathbf{b}^{T}\mathbf{h}(X^{n},U^{l})$
when $U^{l}=\mathbf{S}\circ X^{n}$; see \eqref{eq:def_comb})} 

\For{$i\in[l]$, $j\in[n]$ where $\underline{\mathbf{S}}_{i,j}<\overline{\mathbf{S}}_{i,j}$}

\State{$\hat{\mathbf{S}}\leftarrow\mathbf{S}$}

\State{$\hat{\mathbf{S}}_{i,j}\leftarrow1-\hat{\mathbf{S}}_{i,j}$}\label{step:checkcii2}

\State{$\hat{\mathbf{c}}^{T}\leftarrow\mathbf{b}^{T}\mathrm{sub}(\mathbf{I}_{n};\hat{\mathbf{S}})$} 

\State{\textbf{if} CII $\forall X^{n}:(\mathbf{A}\mathbf{h}(X^{n})\succeq0\to\hat{\mathbf{c}}^{T}\mathbf{h}(X^{n})\ge0)$
}

\State{$\;\;$cannot be verified by \eqref{eq:cii_lp}, \textbf{then}}

\State{$\;\;$set $\underline{\mathbf{S}}_{i,j}\leftarrow\mathbf{S}_{i,j},\;\overline{\mathbf{S}}_{i,j}\leftarrow\mathbf{S}_{i,j}$}

\EndFor

\EndFor

\Until{no further changes are made to $\underline{\mathbf{S}},\overline{\mathbf{S}}$}

\State{\Return $\underline{\mathbf{S}},\overline{\mathbf{S}}$} 

\end{algorithmic}

\caption{\label{alg:sandwich}$\textsc{Sandwich}(\mathbf{A},\mathbf{B},\underline{\mathbf{S}},\overline{\mathbf{S}})$}
\end{algorithm}

Step \ref{step:stildei} sets each auxiliary to the most conservative
choice (set it to the largest if $\mathbf{b}^{T}\mathbf{h}(X^{n},U^{l})$
is increasing, or smallest if decreasing). If $\mathbf{b}^{T}\mathbf{h}(X^{n},\tilde{U}^{l})\ge0$
cannot be verified even for the most conservative choices, then it
cannot possibly be verified by this algorithm. In Step \ref{step:checkcii2},
we set the auxiliaries to the most conservative choice except for
one position. If this change causes $\mathbf{b}^{T}\mathbf{h}(X^{n},\tilde{U}^{l})\ge0$
to fail to be verified, then we know that position must be included
if the change is to exclude it, or that position must be excluded
if the change is to include it, i.e., the choice of whether the position
is included must be fixed to its most conservative choice.
\begin{rem}
We say that $\mathbf{b}^{T}\mathbf{h}(X^{n})$ is \emph{conditionally
increasing} in $X_{i}$ ($i\in[n]$) given $\mathbf{A}\mathbf{h}(X^{n})\succeq0$
if 
\begin{align*}
\forall X^{n},Y:\Big( & \mathbf{A}\mathbf{h}(X^{n})\succeq0\;\wedge\;\mathbf{A}\mathbf{h}\big(X^{i-1},(X_{i},Y),X_{i+1}^{n}\big)\succeq0\\
 & \to\;\mathbf{b}^{T}\mathbf{h}(X^{n})\le\mathbf{b}^{T}\mathbf{h}\big(X^{i-1},(X_{i},Y),X_{i+1}^{n}\big)\Big).
\end{align*}
We may use conditional increasing/decreasing given $\mathbf{A}\mathbf{h}(X^{n})\succeq0$
instead in the sandwich procedure to slightly strengthen the result.
\end{rem}
\medskip{}

\subsection{Cached Linear Program\label{subsec:cachedlp}}

After the sandwich procedure, we have to identify $\mathbf{S}\in\{0,1\}^{l\times n}$
with $\underline{\mathbf{S}}\preceq\mathbf{S}\preceq\overline{\mathbf{S}}$,
such that $\mathbf{A}\mathbf{h}(X^{n})\succeq0\,\rightarrow\,\mathbf{B}\mathbf{h}(X^{n},U^{l})\succeq0$
can be verified when $U^{l}=\mathbf{S}\circ X^{n}$. One method is
to exhaust all $2^{\sum_{i,j}(\overline{\mathbf{S}}_{i,j}-\underline{\mathbf{S}}_{i,j})}$
possible choices of $\mathbf{S}$, and solve the linear program \eqref{eq:cii_lp}
for each combination. We describe a way to skip the linear program
for some of the combinations in Algorithm \ref{alg:cached}. The main
idea is that for $\mathbf{S}\in\{0,1\}^{l\times n}$, if the linear
program \eqref{eq:cii_lp} fails to verify $\mathbf{A}\mathbf{h}(X^{n})\succeq0\,\rightarrow\,\mathbf{B}\mathbf{h}(X^{n},U^{l})\succeq0$
after substituting $U^{l}=\mathbf{S}\circ X^{n}$ (i.e., there exists
a row $\mathbf{b}$ and a vector $\mathbf{k}\in\Gamma_{n}$ that gives
a negative $\mathbf{b}^{T}\mathbf{k}$ in \eqref{eq:cii_lp}; we call
$\mathbf{k}$ a \emph{witness of failure} of $\mathbf{S}$), then
we not only know that $U^{l}=\mathbf{S}\circ X^{n}$ is not a solution
to the EII, but we also guess that the same $\mathbf{k}$ may also
be a witness of failure of another substitution matrix.

\begin{algorithm}[H]
\textbf{$\;\;\;\;$Input:} $\mathbf{A}\in\mathbb{R}^{m_{\mathbf{A}}\times(2^{n}-1)}$,
$\mathbf{B}\in\mathbb{R}^{m_{\mathbf{B}}\times(2^{n+l}-1)}$, lower
and upper bound $\underline{\mathbf{S}},\overline{\mathbf{S}}\in\{0,1\}^{l\times n}$

\textbf{$\;\;\;\;\;\;$} (default values are $\underline{\mathbf{S}}=\mathbf{0}^{l\times n}$,
$\overline{\mathbf{S}}=\mathbf{1}^{l\times n}$)

\textbf{$\;\;\;\;$Output:} $\mathbf{S}\in\{0,1\}^{l\times n}$ or
failure

\smallskip{}

\begin{algorithmic}[1]

\State{$(\underline{\mathbf{S}},\overline{\mathbf{S}})\leftarrow\textsc{Sandwich}(\mathbf{A},\mathbf{B},\underline{\mathbf{S}},\overline{\mathbf{S}})$
(if failure, return failure)}

\State{$\mathcal{K}\leftarrow\emptyset$ (empty list)}

\For{each matrix $\mathbf{S}\in\{0,1\}^{l\times n}$ with $\underline{\mathbf{S}}\preceq\mathbf{S}\preceq\overline{\mathbf{S}}$}

\For{row $\mathbf{b}^{T}$ of $\mathbf{B}$}

\State{$\mathbf{c}^{T}\leftarrow\mathbf{b}^{T}\mathrm{sub}(\mathbf{I}_{n};\mathbf{S})$} 

\State{$\;\;\;\;$(so $\mathbf{c}^{T}\mathbf{h}(X^{n})=\mathbf{b}^{T}\mathbf{h}(X^{n},U^{l})$
when $U^{l}=\mathbf{S}\circ X^{n}$; see \eqref{eq:def_comb})} \label{step:u_subset}

\State{\textbf{if} $\mathbf{c}^{T}\mathbf{k}<0$ for any $\mathbf{k}\in\mathcal{K}$,
} 

\State{$\;\;$\textbf{then} skip $\mathbf{S}$ and continue to the
next matrix } 

\State{Solve the linear program: } 

\State{$\;\;$minimize $\mathbf{c}^{T}\mathbf{k}$ s.t. $\mathbf{k}\in\Gamma_{n}$,
$\mathbf{A}\mathbf{k}\succeq0$, $\mathbf{k}_{2^{n}-1}\le1$} \label{step:lp_bdd}

\State{\textbf{if} optimal value $<0$, \textbf{then} add $\mathbf{k}$
attaining optimum to} 

\State{$\;\;$list $\mathcal{K}$, skip $\mathbf{S}$ and continue
to the next matrix} 

\EndFor

\State{\Return $\mathbf{S}$} 

\EndFor

\State{\Return failure} 

\end{algorithmic}

\caption{\label{alg:cached}$\textsc{VerifyEII}(\mathbf{A},\mathbf{B},\underline{\mathbf{S}},\overline{\mathbf{S}})$}
\end{algorithm}

In Algorithm \ref{alg:cached}, if the linear program gives a negative
optimal value, then the witness of failure $\mathbf{k}$ is stored,
and can be used to quickly check the linear program of another substitution
matrix before running a full linear programming algorithm.  In Step
\ref{step:lp_bdd}, the constraint $\mathbf{k}_{2^{n}-1}\le1$ (meaning
$H(X^{n})\le1$) is to ensure that the linear program is bounded,
so we can obtain the optimal value of $\mathbf{k}$. This caching
optimization gives a significant speedup in the PSITIP implementation.
Note that this optimization can also be applied to the linear programming
steps in the sandwich procedure.

Another optimization is to use the increasing/decreasing information
(computed in the sandwich procedure) to skip some substitution matrices
if a better combination has failed. We will not discuss this optimization
in this paper.

\medskip{}

\section{Inference Rules\label{sec:neii}}

For brevity, we will write the EII in \eqref{eq:aux_impl2} as $\mathbf{A}\stackrel{EII}{\to}\mathbf{B}$.
Note that $n,l$ can be deduced from the widths of $\mathbf{A}$ and
$\mathbf{B}$. In the following sections, we will discuss several
inference rules, which are rules that allow us to deduce new EIIs
from existing EIIs. We first discuss some simple rules, which we call
\emph{elementary rules}. We switch between the notation in \eqref{eq:aux_impl2}
and the ``$\stackrel{EII}{\to}$'' notation depending on which is
simpler. Also recall that the index $i\in\{1,\ldots,2^{n}-1\}$ in
$\mathbf{h}\in\mathbb{R}^{2^{n}-1}$ corresponds to the subset $\mathcal{S}\subseteq[n]$
that is the binary representation of $i$, i.e., $i=\sum_{k\in\mathcal{S}}2^{k-1}$.
\begin{prop}
[Elementary rules]\label{prop:basicrules}For $\mathbf{A}\in\mathbb{R}^{m_{\mathbf{A}}\times(2^{n}-1)}$,
$\mathbf{B}\in\mathbb{R}^{m_{\mathbf{B}}\times(2^{n+l}-1)}$, $\mathbf{C}\in\mathbb{R}^{m_{\mathbf{C}}\times(2^{n+l+k}-1)}$,
$n,l,k\ge0$, we have
\end{prop}
\begin{enumerate}
\item Shannon-type inequality (Sha). $\mathbf{0}^{0\times(2^{3}-1)}\stackrel{EII}{\to}[0,0,0,-1,1,1,-1]$.
Equivalently, $\forall X,Y,Z:\,I(X;Y|Z)\ge0$.
\item Substitution of variables (SubV). For any $n'\in\mathbb{N}$, $\mathbf{S}\in\{0,1\}^{n\times n'}$,
\begin{align*}
\mathbf{A}\stackrel{EII}{\to}\mathbf{B}\,\Rightarrow\, & \mathbf{A}\mathrm{sub}(\mathbf{S})\stackrel{EII}{\to}\mathbf{B}\mathrm{sub}\left(\left[\begin{array}{c|c}
\mathbf{S} & \mathbf{0}^{n\times l}\\
\hline \mathbf{0}^{l\times n'} & \mathbf{I}_{l}
\end{array}\right]\right).
\end{align*}
Equivalently,
\begin{align*}
 & \forall X^{n}:\,\left(\mathbf{A}\mathbf{h}(X^{n})\succeq0\,\to\,\exists U^{l}:\,\mathbf{B}\mathbf{h}(X^{n},U^{l})\succeq0\right)\\
 & \!\!\Rightarrow\,\forall Y^{n'}:\left(\mathbf{A}\mathbf{h}(\mathbf{S}\circ Y^{n'})\succeq0\to\exists U^{l}:\mathbf{B}\mathbf{h}(\mathbf{S}\circ Y^{n'},U^{l})\succeq0\right).
\end{align*}
This corresponds to substituting the variables $X^{n}=\mathbf{S}\circ Y^{n'}$.
\item Substitution of auxiliaries (SubA). For any $l'\in\mathbb{N}$, $\mathbf{S}\in\{0,1\}^{l\times(n+l')}$,
\begin{align*}
 & \mathbf{A}\stackrel{EII}{\to}\mathbf{B}\mathrm{sub}\left(\left[\begin{array}{c}
\mathbf{I}_{n}\,|\,\mathbf{0}^{n\times l'}\\
\hline \mathbf{S}
\end{array}\right]\right)\,\Rightarrow\,\mathbf{A}\stackrel{EII}{\to}\mathbf{B}.
\end{align*}
Equivalently,
\begin{align*}
 & \forall X^{n}:\,\left(\mathbf{A}\mathbf{h}(X^{n})\succeq0\,\to\,\exists V^{l'}:\,\mathbf{B}\mathbf{h}(X^{n},\mathbf{S}\circ(X^{n},V^{l'}))\succeq0\right)\\
 & \!\!\Rightarrow\,\forall X^{n}:\,\left(\mathbf{A}\mathbf{h}(X^{n})\succeq0\,\to\,\exists U^{l}:\,\mathbf{B}\mathbf{h}(X^{n},U^{l})\succeq0\right),
\end{align*}
where $(X^{n},V^{l'})=(X_{1},\ldots,X_{n},V_{1},\ldots,V_{l'})$ is
a random sequence with length $n+l'$. This corresponds to substituting
the auxiliaries $U^{l}=\mathbf{S}\circ(X^{n},V^{l'})$.
\item Conic combination (Cone). Let $\mathbf{D}\in\mathbb{R}_{\ge0}^{\tilde{m}_{\mathbf{A}}\times m_{\mathbf{A}}}$
($\mathbf{D}$ has nonnegative entries). We have
\[
\mathbf{A}\stackrel{EII}{\to}\mathbf{D}\mathbf{A}.
\]
\item Transitivity (Tran). 
\[
\big(\mathbf{A}\stackrel{EII}{\to}\mathbf{B}\;\wedge\;\mathbf{B}\stackrel{EII}{\to}\mathbf{C}\big)\,\Rightarrow\,\mathbf{A}\stackrel{EII}{\to}\mathbf{C}.
\]
Equivalently,
\begin{align*}
 & \forall X^{n}:\,\left(\mathbf{A}\mathbf{h}(X^{n})\succeq0\,\to\,\exists U^{l}:\,\mathbf{B}\mathbf{h}(X^{n},U^{l})\succeq0\right)\\
 & \wedge\,\forall Y^{n+l}:\,\left(\mathbf{B}\mathbf{h}(Y^{n+l})\succeq0\,\to\,\exists V^{k}:\,\mathbf{C}\mathbf{h}(Y^{n+l},V^{k})\succeq0\right)\\
 & \!\!\Rightarrow\,\forall Z^{n}:\left(\mathbf{A}\mathbf{h}(Z^{n})\succeq0\to\exists W^{l+k}:\mathbf{C}\mathbf{h}(X^{n},U^{l},W^{l+k})\succeq0\right),
\end{align*}
which can be seen simply by substituting $X^{n}=Z^{n}$, $Y^{n+l}=(X^{n},U^{l})$
and $W^{l+k}=(U^{l},V^{k})$. Note that transitivity together with
the conic combination rule implies that ``$\stackrel{EII}{\to}$''
is a preorder.
\item Absorption (Abs). 
\[
\mathbf{A}\stackrel{EII}{\to}\mathbf{B}\,\Rightarrow\,\mathbf{A}\stackrel{EII}{\to}\left[\begin{array}{c}
\mathbf{B}\\
\hline \mathbf{A}\,|\,0
\end{array}\right],
\]
where the block matrix on the right hand side is formed by vertically
stacking $\mathbf{B}$ and $\mathbf{A}$ (padded with zeros on the
right to make the width the same as the width of $\mathbf{B}$). Note
that $[\mathbf{A}\,|\,0]\mathbf{h}(X^{n},U^{l})=\mathbf{A}\mathbf{h}(X^{n})$
since $\mathbf{h}(X^{n})$ is the first $2^{n}-1$ entries of $\mathbf{h}(X^{n},U^{l})$.
Equivalently,
\begin{align*}
 & \forall X^{n}:\,\left(\mathbf{A}\mathbf{h}(X^{n})\succeq0\,\to\,\exists U^{l}:\,\mathbf{B}\mathbf{h}(X^{n},U^{l})\succeq0\right)\\
 & \!\!\Rightarrow\,\forall X^{n}:\left(\mathbf{A}\mathbf{h}(X^{n})\succeq0\to\exists U^{l}:\mathbf{B}\mathbf{h}(X^{n},U^{l})\succeq0\,\wedge\,\mathbf{A}\mathbf{h}(X^{n})\succeq0\right).
\end{align*}
\end{enumerate}
We will refer the rules by the abbreviations in parentheses. We remark
that these rules are meant to be computer verifiable, so it is necessary
to specify the rules (even the logically trivial ones) precisely.
We also want to find a minimal set of rules that is sufficient to
prove the results in this paper, so redundant rules (that can be deduced
from other rules) are not included.

For example, the following rule can be deduced from the elementary
rules: For $\mathbf{A}\in\mathbb{R}^{m_{\mathbf{A}}\times(2^{n}-1)}$,
$\mathbf{B}\in\mathbb{R}^{m_{\mathbf{B}}\times(2^{n}-1)}$, $\mathbf{C}\in\mathbb{R}^{m_{\mathbf{C}}\times(2^{n}-1)}$,
we have
\begin{equation}
\big(\mathbf{A}\stackrel{EII}{\to}\mathbf{B}\;\wedge\;\mathbf{A}\stackrel{EII}{\to}\mathbf{C}\big)\,\Rightarrow\,\mathbf{A}\stackrel{EII}{\to}\left[\begin{array}{c}
\mathbf{B}\\
\mathbf{C}
\end{array}\right].\label{eq:rule_concat}
\end{equation}
To show this, 
\begin{align*}
 & \mathbf{A}\stackrel{EII}{\to}\mathbf{B}\;\wedge\;\mathbf{A}\stackrel{EII}{\to}\mathbf{C}\\
 & \Rightarrow\;\mathbf{A}\stackrel{EII}{\to}\mathbf{B}\;\wedge\;\left[\begin{array}{c}
\mathbf{B}\\
\mathbf{A}
\end{array}\right]\stackrel{EII}{\to}\mathbf{A}\;\wedge\;\mathbf{A}\stackrel{EII}{\to}\mathbf{C}\qquad\text{(Cone)}\\
 & \Rightarrow\;\mathbf{A}\stackrel{EII}{\to}\mathbf{B}\;\wedge\;\left[\begin{array}{c}
\mathbf{B}\\
\mathbf{A}
\end{array}\right]\stackrel{EII}{\to}\mathbf{C}\qquad\text{(Tran)}\\
 & \Rightarrow\;\mathbf{A}\stackrel{EII}{\to}\left[\begin{array}{c}
\mathbf{B}\\
\mathbf{A}
\end{array}\right]\;\wedge\;\left[\begin{array}{c}
\mathbf{B}\\
\mathbf{A}
\end{array}\right]\stackrel{EII}{\to}\left[\begin{array}{c}
\mathbf{C}\\
\mathbf{B}\\
\mathbf{A}
\end{array}\right]\qquad\text{(Abs)}\\
 & \Rightarrow\;\mathbf{A}\stackrel{EII}{\to}\left[\begin{array}{c}
\mathbf{C}\\
\mathbf{B}\\
\mathbf{A}
\end{array}\right]\qquad\text{(Tran)}\\
 & \Rightarrow\;\mathbf{A}\stackrel{EII}{\to}\left[\begin{array}{c}
\mathbf{C}\\
\mathbf{B}\\
\mathbf{A}
\end{array}\right]\;\wedge\;\left[\begin{array}{c}
\mathbf{C}\\
\mathbf{B}\\
\mathbf{A}
\end{array}\right]\stackrel{EII}{\to}\left[\begin{array}{c}
\mathbf{B}\\
\mathbf{C}
\end{array}\right]\qquad\text{(Cone)}\\
 & \Rightarrow\;\mathbf{A}\stackrel{EII}{\to}\left[\begin{array}{c}
\mathbf{B}\\
\mathbf{C}
\end{array}\right]\qquad\text{(Tran)}.
\end{align*}

We may regard the elementary rules in Proposition \ref{prop:basicrules}
as the axioms of EIIs. Nevertheless, this set of axioms is not complete.
We will show that the EIIs that can be verified by Algorithm \ref{alg:cached}
are precisely the EIIs that can be deduced using the elementary rules,
i.e., an EII is trivial if and only if it can be deduced using the
elementary rules. The proof of the following theorem is in Appendix
\ref{subsec:pf_trivial}.

\begin{thm}
\label{thm:trivial}For an EII $\mathbf{A}\stackrel{EII}{\to}\mathbf{B}$,
the following are equivalent:
\begin{enumerate}
\item It can be deduced using the elementary rules.
\item $\mathbf{A}\stackrel{EII}{\to}\mathbf{B}$ is trivial \eqref{eq:trivial_imp}
(it can be verified by Algorithm \ref{alg:cached}).
\item There exists $\mathbf{S}\in\{0,1\}^{l\times n}$ and a matrix $\mathbf{D}$
with nonnegative entries such that 
\begin{align*}
 & \mathbf{B}\mathrm{sub}(\mathbf{I}_{n};\mathbf{S})=\mathbf{D}\left[\begin{array}{c}
\mathbf{A}\\
\boldsymbol{\Gamma}_{n}
\end{array}\right],
\end{align*}
where $\boldsymbol{\Gamma}_{n}$ is a matrix such that $\Gamma_{n}=\{\mathbf{k}\in\mathbb{R}^{2^{n}-1}:\,\boldsymbol{\Gamma}_{n}\mathbf{k}\succeq0\}$
(i.e., the rows of $\boldsymbol{\Gamma}_{n}$ corresponds to the elementary
inequalities for Shannon-type inequalities).
\end{enumerate}
\end{thm}

\medskip{}

Note that the only rule that introduces new auxiliary random variables
is the substitution of auxiliaries rule, and hence these rules only
allow us to identify auxiliary random variables that are subsets of
$X^{n}$. The elementary rules alone cannot infer any of the nontrivial
EIIs (e.g. non-Shannon-inequalities and the EIIs in the next section),
and hence they are incomplete if viewed as axioms. The reason we discuss
the elementary rules is that, while they are not powerful enough to
derive nontrivial EIIs on their own, if we assume some nontrivial
EIIs as premises (i.e., introduce more axioms), then the elementary
rules can help us derive more nontrivial EIIs. This will be discussed
in the following sections.

\medskip{}

\section{Nontrivial Existential Information Inequalities\label{sec:nontrivial}}

\subsection{Copy Lemma\label{subsec:copylem}}

We state the copy lemma \cite{zhang1998characterization,dougherty2011non}
as an EII.
\begin{prop}
[Copy lemma \cite{zhang1998characterization,dougherty2011non}]\label{prop:copylem}
For any $n,l\ge0$, 
\begin{align*}
\forall X^{n},Y^{l}:\Big( & \exists U^{l}:\,I(U^{l};Y^{l}|X^{n})=0\\
 & \wedge\;\mathbf{h}(X^{n},U^{l})=\mathbf{h}(X^{n},Y^{l})\Big).
\end{align*}
\end{prop}
Note that we only enforce equalities of entropy terms here instead
of equality in distribution in the original copy lemma, resulting
in a weaker statement.

Even though there are non-Shannon-type inequalities that cannot be
proved using the linear program in \eqref{eq:h2_impl}, all known
unconditional non-Shannon-type inequalities (e.g. \cite{dougherty2006six,xu2008projection,dougherty2011non})
can be proved using the elementary rules and the copy lemma. We remark
that automated application of the copy lemma has been used in \cite{xu2008projection,dougherty2011non}
to discover non-Shannon-type inequalities. The framework in this paper
is more general in the sense that it can incorporate any nontrivial
EII, not only the copy lemma. Refer to Section \ref{sec:auxsearch_premise}
for the algorithm which incorporates known EIIs in the verification
of new EIIs.

The copy lemma can be equivalently stated as the following rule, which
we call the conditional independence rule. Intuitively, the rule states
that if the auxiliaries $U^{l}$ only interact with $X^{n}$, then
$U^{l}$ can be assumed to be conditionally independent of all other
random variables given $X^{n}$. We remark that a similar observation
was made in \cite{kaced2013equivalence}.
\begin{prop}
[Conditional independence rule]\label{prop:cirule} The copy lemma
is equivalent to the following rule:
\begin{align*}
 & \forall X^{n},Y^{k}:\,\Big(\mathbf{A}\mathbf{h}(X^{n},Y^{k})\succeq0\,\to\\
 & \;\;\;\exists U^{l}:\,\mathbf{B}\mathbf{h}(X^{n},U^{l})+\mathbf{C}\mathbf{h}(X^{n},Y^{k})\succeq0\Big)\\
\Rightarrow & \forall X^{n},Y^{k}:\,\Big(\mathbf{A}\mathbf{h}(X^{n},Y^{k})\succeq0\,\to\\
 & \;\;\;\exists U^{l}:\,\mathbf{B}\mathbf{h}(X^{n},U^{l})+\mathbf{C}\mathbf{h}(X^{n},Y^{k})\succeq0\\
 & \;\;\;\;\;\;\wedge\,I(U^{l};Y^{k}|X^{n})=0\Big)
\end{align*}
\end{prop}
\begin{IEEEproof}
Assume the copy lemma is true. Substituting $X^{n}\leftarrow X^{n}$,
$Y^{l}\leftarrow(U^{l},Y)$ to the copy lemma,
\begin{align}
 & \forall X^{n},U^{l},Y:\Big(\exists V^{l},Z:\,I(V^{l},Z;U^{l},Y|X^{n})=0\nonumber \\
 & \;\wedge\;\mathbf{h}(X^{n},V^{l},Z)=\mathbf{h}(X^{n},U^{l},Y)\Big)\nonumber \\
\Rightarrow & \forall X^{n},U^{l},Y:\Big(\mathbf{B}\mathbf{h}(X^{n},U^{l})+\mathbf{C}\mathbf{h}(X^{n},Y^{k})\succeq0\,\to\nonumber \\
 & \;\exists V^{l},Z:\mathbf{B}\mathbf{h}(X^{n},U^{l})+\mathbf{C}\mathbf{h}(X^{n},Y^{k})\succeq0\nonumber \\
 & \;\;\wedge\;I(V^{l},Z;U^{l},Y|X^{n})=0\nonumber \\
 & \;\;\wedge\;\mathbf{h}(X^{n},V^{l},Z)=\mathbf{h}(X^{n},U^{l},Y)\Big),\label{eq:cir_copy}
\end{align}
where the implication is by applying transitivity with $\mathbf{B}\mathbf{h}(X^{n},U^{l})+\mathbf{C}\mathbf{h}(X^{n},Y^{k})\succeq0\,\to\,\mathrm{True}$
(by conic combination), and then by absorption. We have
\begin{align*}
 & \forall X^{n},Y^{k}:\,\Big(\mathbf{A}\mathbf{h}(X^{n},Y^{k})\succeq0\,\to\\
 & \;\;\;\exists U^{l}:\,\mathbf{B}\mathbf{h}(X^{n},U^{l})+\mathbf{C}\mathbf{h}(X^{n},Y^{k})\succeq0\Big)\\
\stackrel{(a)}{\Rightarrow} & \forall X^{n},Y:\Big(\mathbf{A}\mathbf{h}(X^{n},Y^{k})\succeq0\to\\
 & \;\exists U^{l},V^{l},Z:\mathbf{B}\mathbf{h}(X^{n},U^{l})+\mathbf{C}\mathbf{h}(X^{n},Y^{k})\succeq0\\
 & \;\;\wedge\;I(V^{l},Z;U^{l},Y|X^{n})=0\\
 & \;\;\wedge\;\mathbf{h}(X^{n},V^{l},Z)=\mathbf{h}(X^{n},U^{l},Y)\Big)\\
\Rightarrow & \forall X^{n},Y:\Big(\mathbf{A}\mathbf{h}(X^{n},Y^{k})\succeq0\to\\
 & \;\;\;\exists U^{l}:\,\mathbf{B}\mathbf{h}(X^{n},U^{l})+\mathbf{C}\mathbf{h}(X^{n},Y^{k})\succeq0\\
 & \;\;\;\;\;\;\wedge\;I(V^{l};Y|X^{n})=0\Big)\;\;\text{(Sha, Con, Elim)}
\end{align*}
where (a) is by \eqref{eq:cir_copy} and transitivity.

For the other direction, assume the conditional independence rule
is true. We have
\begin{align*}
 & \forall Y^{l}:\left(\exists U^{l}:\,\bigwedge_{i=1}^{l}(H(Y_{i}|U_{i})=H(U_{i}|Y_{i})=0)\right)\;\;\text{(Join, Tran)}\\
\Rightarrow & \forall X^{n},Y^{l}:\left(\exists U^{l}:\,\bigwedge_{i=1}^{l}(H(Y_{i}|U_{i})=H(U_{i}|Y_{i})=0)\right)\;\;\text{(Elim)}\\
\Rightarrow & \forall X^{n},Y^{l}:\left(\exists U^{l}:\,\mathbf{h}(X^{n},U^{l})=\mathbf{h}(X^{n},Y^{l})\right)\;\;\text{(Sha, Con)}\\
\Rightarrow & \forall X^{n},Y^{l}:\Big(\exists U^{l}:\,\mathbf{h}(X^{n},U^{l})=\mathbf{h}(X^{n},Y^{l})\,\wedge\,I(U^{l};Y^{l}|X^{n})=0\Big),
\end{align*}
where the last implication is by the conditional independence rule.
\end{IEEEproof}
\medskip{}

\medskip{}

\subsection{Functional Representation Lemma}

We state the functional representation lemma \cite{elgamal2011network}
as an EII.
\begin{prop}
[Functional representation lemma \cite{elgamal2011network}] 
\[
\forall X,Y:\big(\exists U:\,I(X;U)=H(Y|X,U)=0\big).
\]
\end{prop}
The strong functional representation lemma \cite{sfrl_trans} states
that, in addition to $I(X;U)=H(Y|X,U)=0$, we can also have $H(Y|U)\le I(X;Y)+\log(I(X;Y)+1)+4$.
Technically, the strong functional representation lemma is not an
EII due to the non-linear logarithmic term. Nevertheless, it can still
be handled by the PSITIP implementation using a gap term, and can
be used to prove the achievability of Gelfand-Pinsker theorem \cite{gelfand1980coding}
automatically (see Section \ref{sec:auxsearch_premise}).

\medskip{}

\subsection{Other Examples}

The double Markov property \cite{csiszar2011information} can be stated
as an EII:
\begin{align*}
\forall X,Y,Z:\Big( & I(X;Z|Y)=I(Y;Z|X)=0\;\to\\
 & \exists U:\,H(U|X)=H(U|Y)=I(X,Y;Z|U)=0\Big).
\end{align*}
The existence of the G{\'a}cs-K{\"o}rner common part~\cite{gacs1973common}
can be stated as
\begin{align*}
\forall X,Y:\Big( & \exists U:\,H(U|X)=H(U|Y)=0\\
 & \wedge\;\forall V:\,\big(H(V|X)=H(V|Y)=0\,\to\,H(V|U)=0\big)\Big).
\end{align*}
This can be considered as a ``nested'' EII due to the extra nested
layer of the universally quantified variable $V$.\footnote{A subsequent work by the author \cite{li2021first} studies the first-order
theory of random variables, which allows arbitrary nesting of existential
and universal quantifiers.} Similarly, the existence of the minimal sufficient statistic can
be stated as
\begin{align*}
\forall X,Y:\Big( & \exists U:\,H(U|X)=I(X;Y|U)=0\\
 & \wedge\;\forall V:\,\big(H(V|X)=I(X;Y|V)=0\,\to\,H(U|V)=0\big)\Big).
\end{align*}
We remark that the PSITIP implementation can handle arbitrarily nested
expressions such as the two aforementioned examples.

The infinite divisibility of information \cite{li2020infinite} states
that for any random variable $X$ and positive integer $n$, there
exists i.i.d. $U_{1},\ldots,U_{n}$ such that $H(X|U^{n})=0$ and
$H(U_{1})\le e/(n(e-1))H(X)+2.43$. A weaker form (which does not
require $U_{1},\ldots,U_{n}$ to have the same distribution) can be
stated as
\begin{align*}
\forall X:\Big( & \exists U^{n}:\,H(U_{1})=\cdots=H(U_{n})=\frac{1}{n}H(U^{n})\\
 & \wedge\;H(X|U^{n})=0\;\wedge\;H(U_{1})\le\frac{e}{n(e-1)}H(X)+2.43\Big).
\end{align*}
While this is technically not an EII due to the constant term, it
can still be handled by the PSITIP implementation (which can handle
affine expressions).

\medskip{}

\section{Auxiliary Searching with EII Premises\label{sec:auxsearch_premise}}

Suppose we want to prove that an EII follows from a collection of
EII premises assumed to be true (a premise can be one of the nontrivial
EIIs in Section \ref{sec:nontrivial} that are always true). We describe
how to incorporate the premises using the transitivity rule and the
conditional independence rule (Proposition \ref{prop:cirule}). Suppose
we want to verify the EII \eqref{eq:aux_impl2}: $\forall X^{n}:\,(\mathbf{A}\mathbf{h}(X^{n})\succeq0\,\to\,\exists U^{l}:\,\mathbf{B}\mathbf{h}(X^{n},U^{l})\succeq0)$,
using the premise $\forall Y^{n'}:\,(\mathbf{C}\mathbf{h}(Y^{n'})\succeq0\,\to\,\exists V^{l'}:\,\mathbf{D}\mathbf{h}(Y^{n'},V^{l'})\succeq0)$.
The idea is to apply the premise on $Y^{n'}=\tilde{\mathbf{S}}\circ X^{n}$
where $\tilde{\mathbf{S}}\in\{0,1\}^{n'\times n}$. If we can verify
$\mathbf{C}\mathbf{h}(Y^{n'})\succeq0$ for this choice of $Y^{n'}$,
then we can use the premise to assume that $\exists V^{l'}:\,\mathbf{D}\mathbf{h}(Y^{n'},V^{l'})\succeq0$
is true. We now state this precisely using the elementary rules. Using
the substitution of variables rule to substitute $Y^{n'}=\tilde{\mathbf{S}}\circ X^{n}$
to the premise, and use the substitution of variables rule to introduce
$X_{n'+1}^{n}$, and use the conditional independence rule to obtain
\begin{align}
\forall X^{n} & :\big(\mathbf{C}\mathbf{h}(\tilde{\mathbf{S}}\circ X^{n})\succeq0\to\exists V^{l'}:\,\mathbf{D}\mathbf{h}(\tilde{\mathbf{S}}\circ X^{n},V^{l'})\succeq0\nonumber \\
 & \;\;\;\;\;\;\wedge\;I(V^{l'};X^{n}|\tilde{\mathbf{S}}\circ X^{n})=0\big).\label{eq:prem_ci}
\end{align}
Therefore, if we can verify the CII:
\begin{equation}
\forall X^{n}:\,\big(\mathbf{A}\mathbf{h}(X^{n})\succeq0\,\to\,\mathbf{C}\mathbf{h}(\tilde{\mathbf{S}}\circ X^{n})\succeq0\big),\label{eq:prem_vtop}
\end{equation}
and the EII:
\begin{align}
\forall X^{n} & ,V^{l'}:\big(\mathbf{A}\mathbf{h}(X^{n})\succeq0\;\wedge\;\mathbf{C}\mathbf{h}(\tilde{\mathbf{S}}\circ X^{n})\succeq0\nonumber \\
 & \;\;\;\;\;\wedge\;\mathbf{D}\mathbf{h}(\tilde{\mathbf{S}}\circ X^{n},V^{l'})\succeq0\;\wedge\;I(V^{l'};X^{n}|\tilde{\mathbf{S}}\circ X^{n})=0\nonumber \\
 & \;\;\;\to\;\exists U^{l}:\,\mathbf{B}\mathbf{h}(X^{n},U^{l})\succeq0\big),\label{eq:prem_toshow}
\end{align}
then we can verify the desired EII \eqref{eq:aux_impl2} by applying
the transitivity rule on \eqref{eq:prem_vtop}, \eqref{eq:prem_ci},
\eqref{eq:prem_toshow}. Therefore, to verify the desired EII \eqref{eq:aux_impl2},
we can use the linear program \eqref{eq:cii_lp} to verify \eqref{eq:prem_vtop},
and then the auxiliary searching algorithm (Algorithm \ref{alg:cached})
to verify \eqref{eq:prem_toshow}.

We can exhaust all choices of $\tilde{\mathbf{S}}\in\{0,1\}^{n'\times n}$
and repeat the above procedures. To speed up the search, note that
Algorithm \ref{alg:cached} returns one valid substitution matrix
$\tilde{\mathbf{S}}$ for \eqref{eq:prem_vtop}, and hence can be
modified to enumerate all substitution matrices satisfying \eqref{eq:prem_vtop}
more efficiently. We define a variant of Algorithm \ref{alg:cached},
called $\textsc{VerifyEII\_List}(\mathbf{A},\mathbf{B},\underline{\mathbf{S}},\overline{\mathbf{S}})$,
by making the following change to Algorithm \ref{alg:cached}: 
\begin{itemize}
\item Change the line ``return $\mathbf{S}$'' to ``add $\mathbf{S}$
to a list $\mathcal{T}$ (initialized to $\mathcal{T}=\emptyset$)''.\footnote{A more efficient implementation is to use a generator or coroutine,
e.g. using ``yield $\mathbf{S}$'' in Python, so the function can
stop as soon as one $\mathbf{S}$ gives the desired result.}
\item Change the last line ``return failure'' to ``return $\mathcal{T}$''. 
\end{itemize}
This will allow the modified algorithm to output a list $\mathcal{T}$
of valid substitution matrices $\mathbf{S}$ that proves the EII.

The algorithm for verifying $\mathbf{A}\stackrel{EII}{\to}\mathbf{B}$
under the premise $\mathbf{C}\stackrel{EII}{\to}\mathbf{D}$ is given
in Algorithm \ref{alg:premise}. It returns two matrices $\tilde{\mathbf{S}}\in\{0,1\}^{n'\times n}$,
$\mathbf{S}\in\{0,1\}^{l\times(n+l')}$, which means that the EII
$\mathbf{A}\stackrel{EII}{\to}\mathbf{B}$ can be proved by substituting
$Y^{n'}=\tilde{\mathbf{S}}\circ X^{n}$ and $U^{l}=\mathbf{S}\circ(X^{n},V^{l'})$
(note that $U_{i}$ is allowed to be a combination of the entries
in $X^{n}$ and $V^{l'}$). If the program outputs $\emptyset,\mathbf{S}$,
it means the premise $\mathbf{C}\stackrel{EII}{\to}\mathbf{D}$ is
not needed. We remark that this algorithm is useful not only when
the desired consequence is an EII, but also when it is an UII or CII
(which are special cases of EII). For example, this algorithm can
prove non-Shannon-type UIIs and CIIs using the copy lemma (Proposition
\ref{prop:copylem}) as premise.

\begin{algorithm}[H]
\textbf{$\;\;\;\;$Input:} $\mathbf{A}\in\mathbb{R}^{m_{\mathbf{A}}\times(2^{n}-1)}$,
$\mathbf{B}\in\mathbb{R}^{m_{\mathbf{B}}\times(2^{n+l}-1)}$,

\textbf{$\;\;\;\;$ $\;\;\;\;$$\;\;\;\;$} $\mathbf{C}\in\mathbb{R}^{m_{\mathbf{C}}\times(2^{n'}-1)}$,
$\mathbf{D}\in\mathbb{R}^{m_{\mathbf{D}}\times(2^{n'+l'}-1)}$

\textbf{$\;\;\;\;$Output:} $\tilde{\mathbf{S}}\in\{0,1\}^{n'\times n}$,
$\mathbf{S}\in\{0,1\}^{l\times(n+l')}$, or failure

\smallskip{}

\begin{algorithmic}[1]

\State{$\mathbf{S}\leftarrow\textsc{VerifyEII}(\mathbf{A},\mathbf{B})$}

\State{\textbf{if} not failure \textbf{then} return $\emptyset,\mathbf{S}$} 

\For{$\tilde{\mathbf{S}}\in\textsc{VerifyEII\_List}(\mathbf{A},\mathbf{C})$}

\State{$\mathbf{S}\leftarrow\textsc{VerifyEII}(\mathbf{E},\mathbf{B})$,
where $\mathbf{E}\stackrel{EII}{\to}\mathbf{B}$}\label{step:aux_recur}

\State{$\;\;\;$represents the EII in \eqref{eq:prem_toshow}}

\State{\textbf{if} not failure \textbf{then} return $\tilde{\mathbf{S}},\mathbf{S}$} 

\EndFor

\State{\Return failure}

\end{algorithmic}

\caption{\label{alg:premise}$\textsc{VerifyEII\_Premise}(\mathbf{A},\mathbf{B},\mathbf{C},\mathbf{D})$}
\end{algorithm}

\medskip{}

\begin{rem}
If there are multiple premises, we can  replace line \ref{step:aux_recur}
in Algorithm \ref{alg:premise} by a recursive call to the algorithm
itself, using another premise. Note that even when we only have one
premise, it may be useful to repeat the premise multiple times, so
it can be applied multiple times.
\end{rem}
\medskip{}

\section{Union Rule and Leave-One-Out Procedure}

We present another inference rule, called the union rule.
\begin{prop}
[Union rule] For $\mathbf{A}\in\mathbb{R}^{m_{\mathbf{A}}\times(2^{n}-1)}$,
$\mathbf{B}\in\mathbb{R}^{m_{\mathbf{B}}\times(2^{n+l}-1)}$, $\mathbf{c}\in\mathbb{R}^{2^{n}-1}$,
\[
\left[\begin{array}{c}
\mathbf{A}\\
\mathbf{c}^{T}
\end{array}\right]\stackrel{EII}{\to}\mathbf{B}\;\wedge\;\left[\begin{array}{c}
\mathbf{A}\\
-\mathbf{c}^{T}
\end{array}\right]\stackrel{EII}{\to}\mathbf{B}\;\Rightarrow\;\mathbf{A}\stackrel{EII}{\to}\mathbf{B}.
\]
\end{prop}
We can see that the union rule is true simply by considering the cases
whether $\mathbf{c}^{T}\mathbf{h}(X^{n})\ge0$ (where the result follows
from the first EII $\mathbf{A}\mathbf{h}(X^{n})\succeq0\,\wedge\,\mathbf{c}^{T}\mathbf{h}(X^{n})\ge0\,\to\,\mathbf{B}\mathbf{h}(X^{n})\succeq0$
above) or $\mathbf{c}^{T}\mathbf{h}(X^{n})<0$ (where the result follows
from the second EII). While the union rule is logically trivial, it
is useful since the choices of $U^{l}$ in the two cases do not need
to be the same. We describe how to modify the algorithms in Sections
\ref{sec:auxsearch} and \ref{sec:auxsearch_premise} to automatically
apply the union rule. Suppose we are trying to prove the EII in \eqref{eq:aux_impl2}:
\[
\forall X^{n}:\,\left(\mathbf{A}\mathbf{h}(X^{n})\succeq0\,\to\,\exists U^{l}:\,\mathbf{B}\mathbf{h}(X^{n},U^{l})\succeq0\right),
\]
and we find a substitution $U^{l}=\mathbf{S}\circ X^{n}$ (where $\mathbf{S}\in\{0,1\}^{l\times n}$)
such that $\mathbf{A}\mathbf{h}(X^{n})\succeq0\,\to\,\mathbf{b}^{T}\mathbf{h}(X^{n},\mathbf{S}\circ X^{n})\ge0$
holds for all rows $\mathbf{b}^{T}$ of $\mathbf{B}$ except one,
i.e., it is ``almost correct''. We can let that one row be $\mathbf{b}^{T}$,
$\mathbf{b}\in\mathbb{R}^{2^{n+l}-1}$, and let $\mathbf{c}^{T}=\mathbf{b}^{T}\mathrm{sub}(\mathbf{I}_{n};\mathbf{S})\in\mathbb{R}^{2^{n}-1}$
(such that $\mathbf{c}^{T}\mathbf{h}(X^{n})=\mathbf{b}^{T}\mathbf{h}(X^{n},U^{l})$
for all $X^{n}$ when $U^{l}=\mathbf{S}\circ X^{n}$; see \eqref{eq:def_comb}).
Adding $\mathbf{c}^{T}\mathbf{h}(X^{n})\succeq0$ as an assumption,
we know this EII holds: 
\[
\left[\begin{array}{c}
\mathbf{A}\\
\mathbf{c}^{T}
\end{array}\right]\mathbf{h}(X^{n})\succeq0\,\to\,\exists U^{l}:\,\mathbf{B}\mathbf{h}(X^{n},U^{l})\succeq0.
\]
By the union rule, it is left to prove that 
\[
\left[\begin{array}{c}
\mathbf{A}\\
-\mathbf{c}^{T}
\end{array}\right]\mathbf{h}(X^{n})\succeq0\,\to\,\exists U^{l}:\,\mathbf{B}\mathbf{h}(X^{n},U^{l})\succeq0,
\]
which is easier to prove because of the additional assumption. Therefore,
we can utilize ``almost correct'' choices of $U^{l}$ to strengthen
the assumption. This procedure can be repeated until we find the correct
choice of $U^{l}$. 

Consider the toy example of proving the EII
\[
\forall X,Y:\,\exists U:\,I(X;Y|U)\le0\,\wedge\,2H(U)\le H(X)+H(Y).
\]
Note that there are two rows in $\mathbf{B}$ corresponding to $I(X;Y|U)\le0$
and $2H(U)\le H(X)+H(Y)$. We try different choices of $U$ and note
which of these two rows are satisfied. First we try $U=X$, which
satisfies $I(X;Y|U)\le0$, but does not satisfy $2H(U)\le H(X)+H(Y)$,
which becomes $H(X)\le H(Y)$ after substituting $U=X$. Therefore,
we know that if we add an assumption $H(X)\le H(Y)$, then $U=X$
would be a valid choice. It is left to consider the case $H(X)\ge H(Y)$,
i.e., it is left to prove the EII
\begin{align*}
\forall X,Y: & \big(H(X)\ge H(Y)\,\to\\
 & \exists U:\,I(X;Y|U)\le0\,\wedge\,2H(U)\le H(X)+H(Y)\big).
\end{align*}
We then try $U=Y$. It satisfies both $I(X;Y|U)\le0$ and $2H(U)\le H(X)+H(Y)$
(due to the new assumption $H(X)\ge H(Y)$). Hence $U=Y$ is a valid
choice. The algorithm will conclude that the original EII can be proved
by considering either $U=X$ or $U=Y$.

We call this the \emph{leave-one-out procedure}. Refer to Algorithm
\ref{alg:leaveone} for details. Algorithm \ref{alg:leaveone} combines
the cached linear program in Section \ref{subsec:cachedlp} with the
leave-one-out procedure. For each substitution matrix $\mathbf{S}$,
it stores a list $\mathcal{C}$ of rows $\mathbf{c}$ (refer to the
previous discussion for the meaning of $\mathbf{c}$) where $\mathbf{A}\mathbf{h}(X^{n})\succeq0\,\to\,\mathbf{c}^{T}\mathbf{h}(X^{n})\ge0$
fails. If $\mathcal{C}$ is empty, then we successfully find a valid
substitution matrix $\mathbf{S}$. If $|\mathcal{C}|=1$, then we
can apply the leave-one-out procedure to add the row $-\mathbf{c}^{T}$
to $\mathbf{A}$ (note that the cache $\mathcal{K}$ must be cleared
since $\mathbf{A}\mathbf{k}\succeq0$ for $\mathbf{k}\in\mathcal{K}$
may fail under the new $\mathbf{A}$). Algorithm \ref{alg:leaveone}
outputs a list $\mathcal{T}$ of substitution matrices $\mathbf{S}$
such that the original EII can be proved by considering the cases
$U^{l}=\mathbf{S}\circ X^{n}$ for each $\mathbf{S}\in\mathcal{T}$
(note that this is different from $\textsc{VerifyEII\_List}$ in Section
\ref{sec:auxsearch_premise}, where each $\mathbf{S}$ in the list
$\mathcal{T}$ is sufficient to prove the EII by itself).

Using this procedure, the program can automatically generate proofs
that are impossible using only the elementary rules, e.g. the converse
proof of the capacity region of the state-dependent semideterministic
broadcast channel \cite{lapidoth2012state}.

\begin{algorithm}[H]
\textbf{$\;\;\;\;$Input:} $\mathbf{A}\in\mathbb{R}^{m_{\mathbf{A}}\times(2^{n}-1)}$,
$\mathbf{B}\in\mathbb{R}^{m_{\mathbf{B}}\times(2^{n+l}-1)}$, lower
and upper bound $\underline{\mathbf{S}},\overline{\mathbf{S}}\in\{0,1\}^{l\times n}$

\textbf{$\;\;\;\;\;\;$} (default values are $\underline{\mathbf{S}}=\mathbf{0}^{l\times n}$,
$\overline{\mathbf{S}}=\mathbf{1}^{l\times n}$)

\textbf{$\;\;\;\;$Output:} list $\mathcal{T}$ of substitution matrices
or failure

\smallskip{}

\begin{algorithmic}[1]

\State{$\mathcal{T}\leftarrow\emptyset$ (empty list)}

\State{$(\underline{\mathbf{S}},\overline{\mathbf{S}})\leftarrow\textsc{Sandwich}(\mathbf{A},\mathbf{B},\underline{\mathbf{S}},\overline{\mathbf{S}})$
(if failure, return failure)}

\State{$\mathcal{K}\leftarrow\emptyset$ (empty list)}

\For{each matrix $\mathbf{S}\in\{0,1\}^{l\times n}$ with $\underline{\mathbf{S}}\preceq\mathbf{S}\preceq\overline{\mathbf{S}}$}

\State{$\mathcal{C}\leftarrow\emptyset$ (empty list)}

\For{row $\mathbf{b}^{T}$ of $\mathbf{B}$}

\State{$\mathbf{c}^{T}\leftarrow\mathbf{b}^{T}\mathrm{sub}(\mathbf{I}_{n};\mathbf{S})$} 

\State{$\;\;\;\;$(so $\mathbf{c}^{T}\mathbf{h}(X^{n})=\mathbf{b}^{T}\mathbf{h}(X^{n},U^{l})$
when $U^{l}=\mathbf{S}\circ X^{n}$; see \eqref{eq:def_comb})} 

\If{$\mathbf{c}^{T}\mathbf{k}<0$ for any $\mathbf{k}\in\mathcal{K}$}

\State{Add $\mathbf{c}$ to list $\mathcal{C}$}

\Else

\State{Solve the linear program: }

\State{$\;\;$minimize $\mathbf{c}^{T}\mathbf{k}$ s.t. $\mathbf{k}\in\Gamma_{n}$,
$\mathbf{A}\mathbf{k}\succeq0$, $\mathbf{k}_{2^{n}-1}\le1$} 

\If{optimal value $<0$}

\State{Add $\mathbf{k}$ attaining optimum to list $\mathcal{K}$}

\State{Add $\mathbf{c}$ to list $\mathcal{C}$}

\EndIf

\EndIf

\If{$|\mathcal{C}|\ge2$}

\State{Skip $\mathbf{S}$ and continue to the next matrix }

\EndIf

\EndFor

\If{$\mathcal{C}=\emptyset$}

\State{Add $\mathbf{S}$ to list $\mathcal{T}$}

\State{\Return $\mathcal{T}$}

\ElsIf{$|\mathcal{C}|=1$ (let $\mathcal{C}=\{\mathbf{c}\}$)}

\State{$\mathbf{A}\leftarrow\left[\begin{array}{c}
\mathbf{A}\\
-\mathbf{c}^{T}
\end{array}\right]$}

\State{$\mathcal{K}\leftarrow\emptyset$}

\State{Add $\mathbf{S}$ to list $\mathcal{T}$}

\EndIf

\EndFor

\State{\Return failure}

\end{algorithmic}

\caption{\label{alg:leaveone}$\textsc{VerifyEII\_LeaveOneOut}(\mathbf{A},\mathbf{B},\underline{\mathbf{S}},\overline{\mathbf{S}})$}
\end{algorithm}

\medskip{}

\section{Existential Information Predicate and Simplification of Rate Regions\label{sec:eip}}

An EII is a proposition that is either true of false. We may be interested
in predicates with truth value depending on the distribution of some
random variables. Define the \emph{existential information predicate}
(EIP) on the random sequence $X^{n}$ and inequality matrix $\mathbf{A}\in\mathbb{R}^{m_{\mathbf{A}}\times(2^{n+l}-1)}$
to be the predicate
\[
\exists U^{l}\!:\mathbf{A}\mathbf{h}(X^{n},U^{l})\succeq0.
\]
We denote the above predicate as $\mathrm{EIP}_{\mathbf{A}}(X^{n})$.
Note that $l$ can be deduced from $n$ and the width of $\mathbf{A}$.

Rate regions and bounds in network information theory can often be
stated as EIP. For example, the superposition coding inner bound \cite{bergmans1973random,gallager1974capacity}
for the broadcast channel $p(y_{1},y_{2}|x)$ can be stated as the
following EIP on $X,Y_{1},Y_{2},R_{1},R_{2}$:
\begin{align*}
\exists U:\big( & R_{1}\le I(X;Y_{1}|U)\,\wedge\,R_{2}\le I(U;Y_{2})\\
 & \wedge\,R_{1}+R_{2}\le I(X;Y_{1})\\
 & \wedge\,I(U;Y_{1},Y_{2}|X)=0\big).
\end{align*}
Note that $R_{1},R_{2}$ are real variables (refer to Remark \ref{rem:real}
for how to represent them).

\medskip{}

\subsection{EIP Implication Problem\label{subsec:eip_impl}}

An EII can be stated using EIPs:
\[
(\mathbf{A}\stackrel{EII}{\to}\mathbf{B})\Leftrightarrow\left(\forall X^{n}:\,\mathrm{EIP}_{\mathbf{A}}(X^{n})\to\mathrm{EIP}_{\mathbf{B}}(X^{n})\right).
\]
Note that the $\mathbf{A}$ in the above expression has width $2^{n}-1$,
and hence $\mathrm{EIP}_{\mathbf{A}}(X^{n}):\mathbf{A}\mathbf{h}(X^{n})\succeq0$
does not involve any auxiliary. In general, implication between EIPs
in the form $\forall X^{n}:\,\mathrm{EIP}_{\mathbf{A}}(X^{n})\to\mathrm{EIP}_{\mathbf{B}}(X^{n})$
(where $\mathbf{A}$ can have width larger than $2^{n}-1$) may not
have an equivalent EII, though we have the following sufficient EII:
for any $n,l,k\ge0$, $\mathbf{A}\in\mathbb{R}^{m_{\mathbf{A}}\times(2^{n+l}-1)}$,
$\mathbf{B}\in\mathbb{R}^{m_{\mathbf{A}}\times(2^{n+k}-1)}$, we have
\begin{align}
 & \forall X^{n},U^{l}:\big(\mathbf{A}\mathbf{h}(X^{n},U^{l})\succeq0\to\exists V^{k}:\mathbf{B}\mathbf{h}(X^{n},V^{k})\succeq0\big)\label{eq:eip_impl0}\\
 & \Rightarrow\big(\forall X^{n}:\,\mathrm{EIP}_{\mathbf{A}}(X^{n})\to\mathrm{EIP}_{\mathbf{B}}(X^{n})\big).\label{eq:eip_impl}
\end{align}
Therefore, implications between EIPs may be checked using algorithms
for proving EIIs (applied on the EII in the line \eqref{eq:eip_impl0})
discussed in the previous sections. This is useful for deciding whether
a rate region is included in (i.e., is an inner bound of) another
rate region.

\medskip{}

\subsection{Subtractive EIP Simplification and Inner Bounds\label{subsec:simplify_sub}}

We now discuss methods for simplifying an EIP $\mathrm{EIP}_{\mathbf{A}}(X^{n})$.
Common methods for simplifying a rate region in network information
theory include removing redundant inequalities (i.e., if there is
a row $\mathbf{a}^{T}$ of $\mathbf{A}$ such that $\tilde{\mathbf{A}}\mathbf{h}(X^{n},U^{l})\succeq0\to\mathbf{a}^{T}\mathbf{h}(X^{n},U^{l})\succeq0$,
where $\tilde{\mathbf{A}}$ is $\mathbf{A}$ with row $\mathbf{a}^{T}$
removed, then we can remove row $\mathbf{a}^{T}$ from $\mathbf{A}$)
and Fourier-Motzkin elimination on existentially quantified real variables,
e.g. see \cite{elgamal2011network}. Here we discuss an algorithm,
called the \emph{subtractive simplification procedure}, for reducing
the number of auxiliaries in an EIP.

Suppose we want to know whether $U_{1}$ in the EIP $\mathrm{EIP}_{\mathbf{A}}(X^{n})$:
$\exists U^{l}\!:\mathbf{A}\mathbf{h}(X^{n},U^{l})\succeq0$ can be
removed. If we can find $\mathbf{S}\in\{0,1\}^{l\times(n+l)}$ with
$\mathbf{S}_{i,n+1}=0$ for $i\in[l]$ such that
\begin{align}
\forall X^{n},U^{l}:\big( & \mathbf{A}\mathbf{h}(X^{n},U^{l})\succeq0\,\to\,\nonumber \\
 & \mathbf{A}\mathbf{h}(X^{n},\mathbf{S}\circ(X^{n},U^{l}))\succeq0\big),\label{eq:sim_sub}
\end{align}
then we know that substituting $U^{l}\leftarrow\mathbf{S}\circ(X^{n},U^{l})$
to $\mathrm{EIP}_{\mathbf{A}}(X^{n})$ does not strengthen the EIP.
We also know a priori that substituting a particular choice of auxiliaries
cannot weaken an EIP. Hence if \eqref{eq:sim_sub} holds, then substituting
$U^{l}\leftarrow\mathbf{S}\circ(X^{n},U^{l})$ results in an equivalent
EIP. Since $\mathbf{S}_{i,n+1}=0$ for $i\in[l]$, the resultant EIP
does not contain $U_{1}$, and hence we can remove the auxiliary $U_{1}$. 

To find $\mathbf{S}\in\{0,1\}^{l\times(n+l)}$, we apply the auxiliary
searching algorithm (Algorithm \ref{alg:cached}) on the EII
\begin{align}
\forall X^{n},U^{l}:\big( & \mathbf{A}\mathbf{h}(X^{n},U^{l})\succeq0\,\to\,\nonumber \\
 & \exists V^{l}:\,\mathbf{A}\mathbf{h}(X^{n},V^{l})\succeq0\big)\label{eq:topdown-1}
\end{align}
to search for $\mathbf{S}\in\{0,1\}^{l\times(n+l)}$ with $\mathbf{S}_{i,n+1}=0$
for $i\in[l]$ such that $V^{l}=\mathbf{S}\circ(X^{n},U^{l})$ satisfies
the above EII. Since we want to eliminate $U_{1}$ after substituting
$V_{i}=(X_{\mathcal{S}_{i}},U_{\mathcal{T}_{i}})$, we must have $\mathbf{S}_{i,n+1}=0$
for $i\in[l]$. This can be enforced by removing the index of $U_{1}$
from the initial upper bounds $\overline{\mathbf{S}}$ passed to Algorithm
\ref{alg:cached}.

The \emph{full subtractive simplification procedure} repeats this
step to remove auxiliaries until no more auxiliary can be removed.
Refer to Algorithm \ref{alg:full_sub} for details.

\begin{algorithm}[H]
\textbf{$\;\;\;\;$Input:} $n\in\mathbb{N}$, $\mathbf{A}\in\mathbb{R}^{m_{\mathbf{A}}\times(2^{n+l}-1)}$

\textbf{$\;\;\;\;$Output:} $\mathbf{A}$ after simplification

\smallskip{}

\begin{algorithmic}[1]

\Repeat

\For{$i=1,\ldots,l$ (we try to remove $U_{i}$)}

\State{Let $\mathbf{A}\stackrel{EII}{\to}\mathbf{C}$ be the EII:}

\State{$\;\;\;\;$$\forall X^{n},U^{l}:\big(\mathbf{A}\mathbf{h}(X^{n},U^{l})\succeq0\,\to\,\exists V^{l}:\,\mathbf{A}\mathbf{h}(X^{n},V^{l})\succeq0\big).$}

\State{$\mathbf{S}\leftarrow\textsc{VerifyEII}(\mathbf{A},\mathbf{C},\mathbf{0}^{l\times(n+l)},\overline{\mathbf{S}})$
}

\State{$\;\;\;\;$ where $\overline{\mathbf{S}}_{j,k}=0$ if $k=n+i$
(the index of $U_{i}$), $\overline{\mathbf{S}}_{j,k}=1$ otherwise}

\If{not failure}

\State{Compute $\tilde{\mathbf{A}}\in\mathbb{R}^{m_{\mathbf{A}}\times(2^{n+l-1}-1)}$
such that for any $X^{n},U^{l}$,}

\State{$\;\;\;\;$ $\mathbf{A}\mathbf{h}(X^{n},\mathbf{S}\circ(X^{n},U^{l}))=\tilde{\mathbf{A}}\mathbf{h}(X^{n},U_{[l]\backslash\{i\}})$
}

\State{$\mathbf{A}\leftarrow\tilde{\mathbf{A}}$}

\State{$l\leftarrow l-1$}

\EndIf

\EndFor

\Until{no more auxiliaries are removed}

\State{\Return $\mathbf{A}$}

\end{algorithmic}

\caption{\label{alg:full_sub}$\textsc{Simplify\_Sub\_Full}(n,\mathbf{A})$}
\end{algorithm}

Algorithm \ref{alg:full_sub} can be quite slow since it has to search
for the choices of $l$ auxiliaries $V^{l}$ in the $\textsc{VerifyEII}$
step. To improve the efficiency of the algorithm, we use the observation
that if we are trying to remove the auxiliary $U_{1}$, then we usually
only need to consider the choice of $U_{1}$, that is, we find $\mathbf{S}\in\{0,1\}^{1\times(n+l)}$
with $\mathbf{S}_{1,n+1}=0$ such that it suffices to consider the
choice $U_{1}\leftarrow\mathbf{S}\circ(X^{n},U^{l})$, i.e., 
\begin{align*}
\forall X^{n},U^{l}:\big( & \mathbf{A}\mathbf{h}(X^{n},U^{l})\succeq0\,\to\,\\
 & \mathbf{A}\mathbf{h}(X^{n},\mathbf{S}\circ(X^{n},U^{l}),U_{2}^{l})\succeq0\big).
\end{align*}
For the other auxiliaries $U_{2}^{l}$, we simply leave them unchanged.
If the above CII is true, then we can substitute $U_{1}\leftarrow\mathbf{S}\circ(X^{n},U^{l})$
and remove the auxiliary $U_{1}$. To find $\mathbf{S}$, we apply
Algorithm \ref{alg:cached} on the EII
\begin{align}
\forall X^{n},U^{l}:\big( & \mathbf{A}\mathbf{h}(X^{n},U^{l})\succeq0\,\to\,\nonumber \\
 & \exists V:\,\mathbf{A}\mathbf{h}(X^{n},V,U_{2}^{l})\succeq0\big)\label{eq:topdown}
\end{align}
to search for $\mathbf{S}\in\{0,1\}^{1\times(n+l)}$ with $\mathbf{S}_{1,n+1}=0$
such that $V=\mathbf{S}\circ(X^{n},U^{l})$ satisfies the above EII.
We call this the \emph{partial subtractive simplification procedure},
which is less powerful than the full version, but is significantly
faster since we only search for the choice of one auxiliary $V$.
Refer to Algorithm \ref{alg:full_sub} for details.

\begin{algorithm}[H]
\textbf{$\;\;\;\;$Input:} $n\in\mathbb{N}$, $\mathbf{A}\in\mathbb{R}^{m_{\mathbf{A}}\times(2^{n+l}-1)}$

\textbf{$\;\;\;\;$Output:} $\mathbf{A}$ after simplification

\smallskip{}

\begin{algorithmic}[1]

\Repeat

\For{$i=1,\ldots,l$ (we try to remove $U_{i}$)}

\State{Let $\mathbf{A}\stackrel{EII}{\to}\mathbf{C}$ be the EII:}

\State{$\;\;\;\;$$\forall X^{n},U^{l}:\big(\mathbf{A}\mathbf{h}(X^{n},U^{l})\succeq0\,\to\,\exists V:\,\mathbf{A}\mathbf{h}(X^{n},U^{i-1},V,U_{i+1}^{l})\succeq0\big).$}

\State{$\mathbf{S}\leftarrow\textsc{VerifyEII}(\mathbf{A},\mathbf{C},\mathbf{0}^{1\times(n+l)},\overline{\mathbf{S}})$
}

\State{$\;\;\;\;$ where $\overline{\mathbf{S}}_{1,k}=0$ if $k=n+i$
(the index of $U_{i}$), $\overline{\mathbf{S}}_{1,k}=1$ otherwise}

\If{not failure}

\State{Compute $\tilde{\mathbf{A}}\in\mathbb{R}^{m_{\mathbf{A}}\times(2^{n+l-1}-1)}$
such that for any $X^{n},U^{l}$,}

\State{$\;\;\;\;$ $\mathbf{A}\mathbf{h}(X^{n},U^{i-1},\mathbf{S}\circ(X^{n},U^{l}),U_{i+1}^{l})=\tilde{\mathbf{A}}\mathbf{h}(X^{n},U_{[l]\backslash\{i\}})$
}

\State{$\mathbf{A}\leftarrow\tilde{\mathbf{A}}$}

\State{$l\leftarrow l-1$}

\EndIf

\EndFor

\Until{no more auxiliaries are removed}

\State{\Return $\mathbf{A}$}

\end{algorithmic}

\caption{\label{alg:partial_sub}$\textsc{Simplify\_Sub\_Partial}(n,\mathbf{A})$}
\end{algorithm}

Subtractive simplification is best suited for EIPs with few auxiliaries,
e.g. inner bounds for multiuser coding settings (see Section \ref{sec:dbc}).
It can be inefficient for large number of auxiliaries due to the semi-exhaustive
search for each auxiliary.
\begin{rem}
Note that even if the EII \eqref{eq:topdown} fails for the choice
$V=\mathbf{S}\circ(X^{n},U^{l})$, we can still substitute $U_{1}\leftarrow\mathbf{S}\circ(X^{n},U^{l})$
to the original EIP $\exists U^{l}\!:\mathbf{A}\mathbf{h}(X^{n},U^{l})\succeq0$
in order to obtain a new EIP that implies (i.e., is an inner bound
of) the original EIP.
\end{rem}
\medskip{}

\subsection{Additive EIP Simplification and Outer Bounds\label{subsec:simplify_add}}

Next, we discuss another algorithm, called the \emph{additive simplification
procedure}, for reducing the number of auxiliaries in an EIP. Consider
an EIP $\exists U^{l}\!:\mathbf{A}\mathbf{h}(X^{n},U^{l})\succeq0$.
We want to use it to deduce a new equivalent EIP with auxiliaries
$V^{k}$, $k<l$, where the choice of auxiliaries is $V^{k}=\mathbf{S}\circ(X^{n},U^{l})$
(where $\mathbf{S}\in\{0,1\}^{k\times(n+l)}$). Let 
\[
\mathbf{B}=\mathrm{sub}\left(\left[\begin{array}{c}
\mathbf{I}_{n}\,|\,\mathbf{0}^{n\times l}\\
\hline \mathbf{S}
\end{array}\right]\right)\in\mathbb{R}^{(2^{n+k}-1)\times(2^{n+l}-1)}.
\]
Note that $\mathbf{B}$ satisfies $\mathbf{h}(X^{n},V^{k})=\mathbf{B}\mathbf{h}(X^{n},U^{l})$.
Therefore, the original EIP implies the new EIP
\begin{equation}
\exists V^{k}:\,\mathbf{h}(X^{n},V^{k})\in\mathcal{P},\label{eq:add_relax}
\end{equation}
where 
\[
\mathcal{P}:=\left\{ \mathbf{B}\mathbf{k}:\,\mathbf{k}\in\Gamma_{n+l},\,\mathbf{A}\mathbf{k}\succeq0\right\} .
\]
Note that $\mathcal{P}$ is the polyhedral cone obtained by projecting
the polyhedral cone $\{\mathbf{k}\in\Gamma_{n+l}:\,\mathbf{A}\mathbf{k}\succeq0\}$
by the matrix $\mathbf{B}$. It can be computed using any algorithm
for polyhedron projection, for example, Fourier-Motzkin elimination,
or the convex hull method \cite{lassez1992quantifier} (which is used
in the PSITIP implementation). We note that the convex hull method
is also used in \cite{xu2008projection} for the discovery of non-Shannon-type
inequalities, though our framework is more general since it is capable
of incorporating any EII, not only the copy lemma.

To check whether \eqref{eq:add_relax} is equivalent to the original
EIP, it remains to check the other direction of the implication, which
can be checked using the method in Section \ref{subsec:eip_impl}.
If \eqref{eq:add_relax} is equivalent to the original EIP, then we
can simplify the original EIP to \eqref{eq:add_relax}, which contains
fewer auxiliaries since $k<l$. The \emph{additive simplification
procedure} simply perform this checking for every substitution matrix
$\mathbf{S}\in\{0,1\}^{k\times(n+l)}$. Refer to Algorithm \ref{alg:add_sim}
for details.

\begin{algorithm}[H]
\textbf{$\;\;\;\;$Input:} $n\in\mathbb{N}$, $\mathbf{A}\in\mathbb{R}^{m_{\mathbf{A}}\times(2^{n+l}-1)}$

\textbf{$\;\;\;\;$Output:} $\mathbf{A}$ after simplification

\smallskip{}

\begin{algorithmic}[1]

\For{$k=0,1,\ldots,l-1$}

\For{each matrix $\mathbf{S}\in\{0,1\}^{k\times(n+l)}$}

\State{$\mathbf{B}\leftarrow\mathrm{sub}\left(\left[\begin{array}{c}
\mathbf{I}_{n}\,|\,\mathbf{0}^{n\times l}\\
\hline \mathbf{S}
\end{array}\right]\right)$}

\State{Compute the facets of the projected cone:}

\State{$\;\;\;\;$$\mathcal{P}\leftarrow\{\mathbf{B}\mathbf{k}:\,\mathbf{k}\in\Gamma_{n+l},\,\mathbf{A}\mathbf{k}\succeq0\}$}

\State{Compute $\tilde{\mathbf{A}}$ such that $\mathbf{h}(X^{n},V^{k})\in\mathcal{P}\,\Leftrightarrow\,\tilde{\mathbf{A}}\mathbf{h}(X^{n},V^{k})\succeq0$}

\State{Apply $\textsc{VerifyEII}$ (or $\textsc{VerifyEII\_LeaveOneOut}$)
to check the EII:}\label{step:add_sim_verify}

\State{$\;\;\;$$\forall X^{n},V^{k}:\big(\tilde{\mathbf{A}}\mathbf{h}(X^{n},V^{k})\succeq0\to\exists U^{l}:\mathbf{A}\mathbf{h}(X^{n},U^{l})\succeq0\big)$}

\If{not failure}

\State{\Return $\tilde{\mathbf{A}}$}

\EndIf

\EndFor

\EndFor

\State{\Return $\mathbf{A}$}

\end{algorithmic}

\caption{\label{alg:add_sim}$\textsc{Simplify\_Add}(n,\mathbf{A})$}
\end{algorithm}

If we are only interested in obtaining an outer bound of the EIP,
then it is unnecessary to check that the new EIP implies the original
EIP (step \ref{step:add_sim_verify} in Algorithm \ref{alg:full_sub}).
Each $\mathbf{S}$ produces an outer bound $\exists V^{k}:\,\mathbf{h}(X^{n},V^{k})\in\mathcal{P}$
of the original EIP. We then choose the tightest outer bound by comparing
these outer bounds using the method in Section \ref{subsec:eip_impl}.
Given the desired number of auxiliaries $k$, Algorithm \ref{alg:add_outer},
which we call the \emph{additive outer bound procedure}, finds an
outer bound of the given EIP with $k$ auxiliaries.

\begin{algorithm}[H]
\textbf{$\;\;\;\;$Input:} $n\in\mathbb{N}$, $\mathbf{A}\in\mathbb{R}^{m_{\mathbf{A}}\times(2^{n+l}-1)}$,
target number of auxiliaries $k$

\textbf{$\;\;\;\;$Output:} $\mathbf{C}$ such that $\mathrm{EIP}_{\mathbf{C}}(X^{n})$
is an outer bound of $\mathrm{EIP}_{\mathbf{A}}(X^{n})$

\smallskip{}

\begin{algorithmic}[1]

\State{$\mathbf{C}\leftarrow\emptyset$}

\For{each matrix $\mathbf{S}\in\{0,1\}^{k\times(n+l)}$}

\State{$\mathbf{B}\leftarrow\mathrm{sub}\left(\left[\begin{array}{c}
\mathbf{I}_{n}\,|\,\mathbf{0}^{n\times l}\\
\hline \mathbf{S}
\end{array}\right]\right)$}

\State{Compute the facets of the projected cone:}

\State{$\;\;\;\;$$\mathcal{P}\leftarrow\{\mathbf{B}\mathbf{k}:\,\mathbf{k}\in\Gamma_{n+l},\,\mathbf{A}\mathbf{k}\succeq0\}$}

\State{Compute $\tilde{\mathbf{A}}$ such that $\mathbf{h}(X^{n},V^{k})\in\mathcal{P}\,\Leftrightarrow\,\tilde{\mathbf{A}}\mathbf{h}(X^{n},V^{k})\succeq0$}

\If{$\mathbf{C}=\emptyset$}

\State{$\mathbf{C}\leftarrow\tilde{\mathbf{A}}$}

\Else

\State{Apply $\textsc{VerifyEII}$ (or $\textsc{VerifyEII\_LeaveOneOut}$)
to check the EII:}\label{step:add_sim_verify-1}

\State{$\;\;\;$$\forall X^{n},U^{k}:\big(\mathbf{C}\mathbf{h}(X^{n},U^{k})\succeq0\to\exists V^{k}:\tilde{\mathbf{A}}\mathbf{h}(X^{n},V^{k})\succeq0\big)$}

\If{failure}

\State{$\mathbf{C}\leftarrow\tilde{\mathbf{A}}$}

\EndIf

\EndIf

\EndFor

\State{\Return $\mathbf{C}$}

\end{algorithmic}

\caption{\label{alg:add_outer}$\textsc{Outer\_Add}(n,\mathbf{A},k)$}
\end{algorithm}

The additive outer bound procedure is suitable when the number of
auxiliaries is large, and only an outer bound is desired. This is
useful for automated discovery of outer bounds for multiuser coding
settings, where we may encounter an outer bound with too many auxiliaries,
and we are interested in a (possibly weaker) outer bound with fewer
auxiliaries (see Section \ref{sec:dbc} for an example). This is also
useful for discovering non-Shannon inequalities (see Section \ref{sec:outer}
for an example).

All the aforementioned methods (removing redundant inequalities, Fourier-Motzkin
elimination, removing redundant auxiliaries) are implemented in the
PSITIP implementation. The PSITIP implementation allows automated
discovery of inner and outer bounds, using only the graphical representation
of the network as input, via the inner bound in \cite{lee2015unified},
Gallager's approach \cite{gallager1974capacity} for outer bounds,
together with the aforementioned simplification procedures.

\medskip{}

\begin{rem}
The simplification methods can sometimes eliminate the need of auxiliary
random variables and convert an EIP to a non-existential information
predicate (e.g. $\exists U:I(X;U)=I(X;Y|U)=0$ can be simplified to
$I(X;Y)=0$ by identifying $U=Y$). This is known as quantifier elimination
in mathematical logic. Examples of quantifier elimination includes
Fourier-Motzkin elimination and the Tarski-Seidenberg theorem \cite{tarski1951decision,seidenberg1954new}
for real numbers. While the Tarski-Seidenberg theorem establishes
the decidability of the theory of real closed fields by showing that
quantifier elimination is always possible, the same is not true for
information inequalities. For example, the EIP $\exists U:(H(U|X)=0\,\wedge\,H(U)=H(X)/2)$
cannot be reduced to any statement that only concerns $H(X)$ (i.e.,
for any $t>0$, we can construct $X$ where $H(X)=t$ and the EIP
holds, and also construct $X$ where $H(X)=t$ and the EIP does not
hold), and hence the auxiliary $U$ cannot be eliminated. Note that
CIIs, and hence EIIs, are undecidable \cite{li2022ncundecidability}.
\end{rem}

\medskip{}

\section{Examples}

\subsection{Degraded Broacast Channel\label{sec:dbc}}

In this section, we demonstrate the use of the algorithm in finding
and proving the capacity region of 2-receiver degraded broadcast channel
\cite{bergmans1973random,gallager1974capacity}. The Python code for
finding the capacity region (which uses the PSITIP package that implements
the algorithms in this paper,\footnote{Source code of Python Symbolic Information Theoretic Inequality Prover
(PSITIP) is available at \href{https://github.com/cheuktingli/psitip}{https://github.com/cheuktingli/psitip}} and Pyomo \cite{hart2017pyomo} and GLPK \cite{makhorin2008glpk}
for linear programming) is given below:

{\small

\begin{lstlisting}
from psitip import *
PsiOpts.setting(solver = "pyomo.glpk")
PsiOpts.setting(str_style = "latex")

X, Y, Z = rv("X, Y, Z")
M1, M2 = rv_array("M", 1, 3)
R1, R2 = real_array("R", 1, 3)

model = CodingModel()
model.set_rate(M1, R1)    # Rate of M1 is R1
model.set_rate(M2, R2)    # Rate of M2 is R2
model.add_node(M1+M2, X)  # Encoder maps M1,M2 to X
model.add_edge(X, Y)      # Channel X -> Y -> Z
model.add_edge(Y, Z)
model.add_node(Y, M1)     # Decoder1 maps Y to M1
model.add_node(Z, M2)     # Decoder2 maps Z to M2

# Inner bound via [Lee-Chung 2015] without simplification
r_lc = model.get_inner(skip_simplify = True)
print(r_lc)

# Turn on simplification, recover superposition region
r = model.get_inner()
print(r)

# Automatic outer bound without simplification
r_out = model.get_outer()
print(r_out)
# Converse proof, print auxiliary random variables
print((r_out >> r).check_getaux())

# Automatic outer bound, with at most 1 auxiliary
r_out2 = model.get_outer(1)
print(r_out2)
\end{lstlisting}

}

Note that the user only needs to specify the graphical representation
of the setting in the code. The first output of the program (inner
bound via \cite{lee2015unified} without simplification) is given
below (the program directly outputs LaTeX code):

\begin{equation}
\exists A_{M_{1}},A_{M_{2}}:\,\left\{ \begin{array}{l}
\left\{ \begin{array}{l}
R_{1}\ge0,\\
R_{2}\ge0,\\
R_{1}\le I(A_{M_{1}};Y),\\
R_{2}\le I(A_{M_{2}};Z),\\
R_{1}+R_{2}\le I(A_{M_{1}};Y)+I(A_{M_{2}};Z)-I(A_{M_{1}};A_{M_{2}}),\\
(A_{M_{1}},A_{M_{2}})\leftrightarrow X\leftrightarrow Y\leftrightarrow Z
\end{array}\right\} \\
\vee\\
\left\{ \begin{array}{l}
R_{1}\ge0,\\
R_{2}\ge0,\\
R_{1}\le I(A_{M_{1}};Y),\\
R_{2}\le I(A_{M_{2}};A_{M_{1}},Z),\\
R_{1}+R_{2}\le I(A_{M_{1}},A_{M_{2}};Z),\\
(A_{M_{1}},A_{M_{2}})\leftrightarrow X\leftrightarrow Y\leftrightarrow Z
\end{array}\right\} \\
\vee\\
\left\{ \begin{array}{l}
R_{1}\ge0,\\
R_{2}\ge0,\\
R_{2}\le I(A_{M_{2}};Z),\\
R_{1}\le I(A_{M_{1}};A_{M_{2}},Y),\\
R_{1}+R_{2}\le I(A_{M_{2}};Z)+I(A_{M_{1}};Y|A_{M_{2}}),\\
(A_{M_{1}},A_{M_{2}})\leftrightarrow X\leftrightarrow Y\leftrightarrow Z
\end{array}\right\} \\
\vee\\
\left\{ \begin{array}{l}
R_{1}\ge0,\\
R_{2}\ge0,\\
R_{1}\le I(A_{M_{1}};A_{M_{2}},Y),\\
R_{2}\le I(A_{M_{2}};A_{M_{1}},Z),\\
R_{1}+R_{2}\le I(A_{M_{1}},A_{M_{2}};Z),\\
(A_{M_{1}},A_{M_{2}})\leftrightarrow X\leftrightarrow Y\leftrightarrow Z
\end{array}\right\} 
\end{array}\right\} .\label{eq:out1}
\end{equation}
This is the union of 4 EIPs (separated by ``$\vee$'' which means
``or''). Note that $A_{M_{1}}$ and $A_{M_{2}}$ are auxiliary random
variables, and ``$\leftrightarrow$'' denotes Markov chain. Each
individual EIP is obtained via a different choice of simultaneous
nonunique decoding sets in \cite{lee2015unified}. While the inner
bound in \cite{lee2015unified} is general, it requires many manual
choices of parameters in the coding scheme (e.g. the simultaneous
nonunique decoding sets), and hence requires effort to compare the
regions obtained with different choices of parameters in order to
find the largest.

The second output of the program (inner bound with simplification
turned on) is given below:
\begin{equation}
\exists A_{M_{2}}:\,\left\{ \begin{array}{l}
R_{1}\ge0,\\
R_{2}\ge0,\\
R_{2}\le I(A_{M_{2}};Z),\\
R_{1}+R_{2}\le I(A_{M_{2}};Z)+I(X;Y|A_{M_{2}}),\\
A_{M_{2}}\leftrightarrow X\leftrightarrow Y\leftrightarrow Z
\end{array}\right\} .\label{eq:out2}
\end{equation}
This EIP is the same as the superposition region \cite{bergmans1973random,gallager1974capacity}.
The algorithm simplifies the union in \eqref{eq:out1} into the EIP
in \eqref{eq:out2} by comparing the 4 EIPs in \eqref{eq:out1} using
\eqref{eq:eip_impl} and the auxiliary searching algorithm (Algorithm
\ref{alg:cached}) to find the largest, and using the subtractive
simplification procedure described in Section \ref{subsec:simplify_sub}
to reduce the number of auxiliaries (the auxiliary $A_{M_{1}}$ is
removed). We can see how the algorithm eliminates the need of manual
efforts to compare and simplify rate regions.

The third output of the program (automatic outer bound) is an EIP
obtained via the Bayesian network of the past, present and future
random variables (e.g. for $Y^{n}$, the past is $Y_{P}:=Y_{1}^{Q-1}$,
the present is $Y_{Q}$, and the future is $Y_{F}:=Y_{Q+1}^{n}$,
where $Q\sim\mathrm{Unif}[n]$), and the Csisz{\'a}r sum identity
\cite{korner1977images,csiszar1978broadcast} applied on every triple
of random variables:
\begin{align}
 & \exists Y_{P},Y_{F},Z_{P},M_{1},M_{2}:\,\nonumber \\
 & \left\{ \begin{array}{l}
I(M_{1};Y;M_{2}|Y_{P})\le0,\\
R_{1}\le I(M_{1};Y|Y_{P}),\\
I(M_{2};Z;M_{1}|Z_{P})\le0,\\
R_{2}\le I(M_{2};Z|Z_{P}),\\
R_{1}\ge0,\\
R_{2}\ge0,\\
I(Y_{F};Y|Y_{P})=I(Y_{P};Y|Y_{F}),\\
I(Y_{F};Y|Y_{P},M_{1})=I(Y_{P};Y|Y_{F},M_{1}),\\
I(Y_{F};Y|Y_{P},M_{2})=I(Y_{P};Y|Y_{F},M_{2}),\\
I(Y_{F};Y|Y_{P},M_{1},M_{2})=I(Y_{P};Y|Y_{F},M_{1},M_{2}),\\
I(Y_{F};Z|Z_{P})=I(Z_{P};Y|Y_{F}),\\
I(Y_{F};Z|Z_{P},M_{1})=I(Z_{P};Y|Y_{F},M_{1}),\\
I(Y_{F};Z|Z_{P},M_{2})=I(Z_{P};Y|Y_{F},M_{2}),\\
I(Y_{F};Z|Z_{P},M_{1},M_{2})=I(Z_{P};Y|Y_{F},M_{1},M_{2}),\\
I(Y_{F};Y,Z|Y_{P},Z_{P})=I(Y_{P},Z_{P};Y|Y_{F}),\\
I(Y_{F};Y,Z|Y_{P},Z_{P},M_{1})=I(Y_{P},Z_{P};Y|Y_{F},M_{1}),\\
I(Y_{F};Y,Z|Y_{P},Z_{P},M_{2})=I(Y_{P},Z_{P};Y|Y_{F},M_{2}),\\
I(Y_{F};Y,Z|Y_{P},Z_{P},M_{1},M_{2})=I(Y_{P},Z_{P};Y|Y_{F},M_{1},M_{2}),\\
H(X|M_{1},M_{2},Y_{P})=0,\\
H(X|M_{1},M_{2},Y_{F})=0,\\
H(X|M_{1},M_{2},Z_{P})=0,\\
M_{1}{\perp\!\!\perp}M_{2},\\
(M_{1},M_{2},Y_{F})\leftrightarrow X\leftrightarrow Y\leftrightarrow Z,\\
(Y,Z)\leftrightarrow(M_{1},M_{2},Y_{F},X)\leftrightarrow Y_{P}\leftrightarrow Z_{P}
\end{array}\right\} .\label{eq:out3}
\end{align}
 We write $I(X;Y;Z)=I(X;Y)-I(X;Y|Z)$. Note that the time-sharing
random variable $Q$ is absorbed into the auxiliaries and is not present.
Also, the program omits some past/future random variables (e.g. $X_{P},X_{F},Z_{F}$)
that are determined to be unnecessary.\footnote{Technically, instead of $R_{1}\le I(M_{1};Y|Y_{P})$, we should have
$R_{1}\le I(M_{1};Y|Y_{P})+\epsilon$ where $\epsilon\to0$ as the
error probability tends to $0$, though this is unnecessary since
$\epsilon$ can be absorbed into $R_{1}$.} Next, the program checks the optimality of the inner bound by showing
that the outer bound implies the inner bound. The program uses \eqref{eq:eip_impl}
and Algorithm \ref{alg:cached} to show that \eqref{eq:out3} implies
\eqref{eq:out2}. It outputs the choice of auxiliary $A_{M_{2}}=(Y_{P},M_{2})$.

Finally, the program uses the additive outer bound procedure in Section
\ref{subsec:simplify_add} on the bound in \eqref{eq:out3} to find
an outer bound with one auxiliary. It outputs the superposition region
\cite{bergmans1973random,gallager1974capacity}:
\[
\exists A:\,\left\{ \begin{array}{l}
R_{1}\ge0,\\
R_{2}\ge0,\\
R_{2}\le I(A;Z),\\
R_{1}\le I(X;Y|A),\\
A\leftrightarrow X\leftrightarrow Y\leftrightarrow Z
\end{array}\right\} .
\]
We can see that for both automated inner bound and outer bound, the
program correctly outputs the superposition region.

\medskip{}

\subsection{Non-Shannon Inequalities\label{sec:outer}}

To demonstrate the additive outer bound procedure in Section \ref{subsec:simplify_add},
we use it to discover non-Shannon-type inequalities in a way similar
to \cite{xu2008projection}. Python code:\footnote{This code is also given as an example in \href{https://github.com/cheuktingli/psitip}{https://github.com/cheuktingli/psitip}}

{\small

\begin{lstlisting}
from psitip import *
PsiOpts.setting(solver = "pyomo.glpk")  # Set linear programming solver
X, Y, Z, W, U = rv("X, Y, Z, W, U")     # Declare random variables

# State the copy lemma as an EII
r = eqdist([X, Y, U], [X, Y, Z]).exists(U)

# Automatically discover non-Shannon-type inequalities using copy lemma
print(r.discover([X, Y, Z, W]).simplified().latex())
\end{lstlisting}

}

Output of the program, which contains the Zhang-Yeung inequality \cite{zhang1998characterization}
(note that $I(X;Y;Z)=I(X;Y)-I(X;Y|Z)$):

\[
\left\{ \begin{array}{l}
I(X;W;Z)\le I(Y;W)+I(Z;Y|X)+2I(Z;X|Y)+I(X;Y|Z),\\
I(Z;Y;X)\le I(Z;W)+I(Y;Z|X)+I(X;Y|Z)+I(Z;X|Y)+I(Y;X|W),\\
I(Y;W;Z)+I(X;Z;Y)\le I(X;W)+I(Y;Z|X)+I(Z;X|Y)+I(Y;X,Z)
\end{array}\right\} .
\]

\[
\]

\[
\]

\medskip{}

\section{List of Results Provable by the Framework\label{sec:list}}

The following is a non-exhaustive list of results that can be proved
by the PSITIP implementation of the framework.\footnote{Source code is available at \href{https://github.com/cheuktingli/psitip}{https://github.com/cheuktingli/psitip}
. Jupyter notebooks of the proofs are available at \href{https://nbviewer.jupyter.org/github/cheuktingli/psitip/tree/master/examples/}{https://nbviewer.jupyter.org/github/cheuktingli/psitip/tree/master/examples/}
.} Achievability results are obtained via the inner bound in \cite{lee2015unified}
together with the simplification procedures in this paper. In some
cases marked with ``(simplest)'', the framework can automatically
derive the inner bound in its simplest form using the graphical representation
of the network as input. For inner and outer bound results not marked
with ``(simplest)'', the framework can only derive a more complicated
bound, and verify that it is at least as good as the known bound (and
hence give an automated proof of the known bound). 

\begin{enumerate}
\item Gelfand-Pinsker theorem \cite{gelfand1980coding} for channels with
state, achievability (simplest) and converse.
\item Shannon's result \cite{shannon1958channels} on the capacity of channels
with state available causally at the encoder, achievability (simplest)
and converse.
\item 2-receiver degraded broadcast channel \cite{bergmans1973random,gallager1974capacity},
achievability (simplest) and converse.
\item 2-receiver broadcast channel: Marton's inner bound \cite{marton1979broadcast,gelfand1980capacity,liang2007broadcast}
(simplest) and Nair-El Gamal outer bound \cite{nair2006outer}.
\item 2-receiver broadcast channel with less noisy and more capable \cite{korner1977comparison,elgamal1979broadcast}
receivers, achievability (simplest) and converse.
\item State-dependent semideterministic broadcast channel \cite{lapidoth2012state},
achievability (simplest) and converse of capacity region.
\item Multiple access channel \cite{ahlswede1971multi,liao1972multiple,ahlswede1974capacity},
achievability (simplest) and converse of capacity region.
\item Interference channel, Han-Kobayashi inner bound \cite{han1981new}.
\item Wyner-Ahlswede-K{ö}rner network \cite{wyner1975source,ahlswede1975source},
achievability (simplest) and converse of capacity region.
\item Wyner-Ziv theorem \cite{wyner1976ratedistort} for lossy source coding
with side information, achievability (simplest) and converse.
\item Gray-Wyner network \cite{gray1974source}, achievability and converse.
\item Slepian-Wolf theorem \cite{slepianwolf1973a} for distributed lossless
source coding, achievability (simplest) and converse.
\item Distributed lossy source coding: Berger-Tung inner bound \cite{berger1978multiterminal,tung1978multiterminal}
(simplest) and Berger-Tung outer bound \cite{berger1978multiterminal,tung1978multiterminal}.
\item Multiple description coding: Zhang-Berger inner bound \cite{zhang1987new}
(simplest), Venkataramani-Kramer-Goyal inner bound \cite{venkataramani2003multiple,wang2011role}.
\item Vámos network \cite{dougherty2007networks}: upper bound $10/11$
of capacity in \cite{dougherty2007networks} via Zhang-Yeung inequality
\cite{zhang1998characterization}, upper bound $5/6$ of linear coding
capacity in \cite{dougherty2007networks} via Ingleton inequality
\cite{ingleton1971representation}.
\item Wyner's common information \cite{wyner1975common}: data processing
property, lower-bounded by $I(X;Y)$, tensorization.
\item G{\'a}cs-K{\"o}rner common information \cite{gacs1973common}: data
processing property, upper-bounded by $I(X;Y)$, tensorization.
\item Common entropy \cite{kumar2014exact}, data processing property, lower-bounded
by Wyner's common information, subadditivity.
\item Excess functional information \cite{sfrl_trans}, upper bounds in
\cite[Proposition 3.5]{sfrl_trans}.
\item Non-Shannon-type inequalities such as the Zhang-Yeung inequality \cite{zhang1998characterization}
and \cite{dougherty2006six}. 
\end{enumerate}

\medskip{}

\section{Acknowledgement}

This work was supported in part by the Hong Kong Research Grant Council
Grant ECS No. CUHK 24205621, and the Direct Grant for Research, The
Chinese University of Hong Kong (Project ID: 4055133). The author
would like to thank Raymond W. Yeung, Chandra Nair, Laigang Guo, Amos
Lapidoth, Ligong Wang, Si-Hyeon Lee and Sae-Young Chung for their
invaluable comments.

\medskip{}

\appendix
\[
\]

\subsection{Proof of Theorem \ref{thm:trivial} \label{subsec:pf_trivial}}

The equivalence between statement 2 and statement 3 follows directly
from the duality of linear programming.

We then prove if $\mathbf{A}\stackrel{EII}{\to}\mathbf{B}$ satisfies
statement 3, then it can be deduced using the elementary rules. Assume
statement 3 is satisfied. By (Sha), $\forall X,Y,Z:\,I(X;Y|Z)\ge0$.
By (SubV), $\forall X^{n}:\,I(X_{\mathcal{A}};X_{\mathcal{B}}|X_{\mathcal{C}})\ge0$
for any $\mathcal{A},\mathcal{B},\mathcal{C}\subseteq[n]$. Applying
\eqref{eq:rule_concat} repeatedly, we have $\forall X^{n}:\,\boldsymbol{\Gamma}_{n}\mathbf{h}(X^{n})\succeq0$,
or equivalently, $\mathbf{0}^{0\times(2^{n}-1)}\stackrel{EII}{\to}\boldsymbol{\Gamma}_{n}$.
We have
\begin{align*}
 & \mathbf{A}\stackrel{EII}{\to}\mathbf{0}^{0\times(2^{n}-1)}\;\wedge\;\mathbf{0}^{0\times(2^{n}-1)}\stackrel{EII}{\to}\boldsymbol{\Gamma}_{n}\qquad\text{(Cone)}\\
 & \Rightarrow\;\mathbf{A}\stackrel{EII}{\to}\boldsymbol{\Gamma}_{n}\qquad\text{(Tran)}\\
 & \Rightarrow\;\mathbf{A}\stackrel{EII}{\to}\left[\begin{array}{c}
\boldsymbol{\Gamma}_{n}\\
\mathbf{A}
\end{array}\right]\qquad\text{(Abs)}\\
 & \Rightarrow\;\mathbf{A}\stackrel{EII}{\to}\left[\begin{array}{c}
\boldsymbol{\Gamma}_{n}\\
\mathbf{A}
\end{array}\right]\;\wedge\;\left[\begin{array}{c}
\boldsymbol{\Gamma}_{n}\\
\mathbf{A}
\end{array}\right]\stackrel{EII}{\to}\mathbf{D}\left[\begin{array}{c}
\mathbf{A}\\
\boldsymbol{\Gamma}_{n}
\end{array}\right]\qquad\text{(Cone)}\\
 & \Rightarrow\;\mathbf{A}\stackrel{EII}{\to}\mathbf{D}\left[\begin{array}{c}
\mathbf{A}\\
\boldsymbol{\Gamma}_{n}
\end{array}\right]\qquad\text{(Tran)}\\
 & \Rightarrow\;\mathbf{A}\stackrel{EII}{\to}\mathbf{B}\mathrm{sub}(\mathbf{I}_{n};\mathbf{S})\\
 & \Rightarrow\,\mathbf{A}\stackrel{EII}{\to}\mathbf{B}\qquad\text{(SubA)}.
\end{align*}

Finally prove that if $\mathbf{A}\stackrel{EII}{\to}\mathbf{B}$ can
be deduced using the elementary rules, then it satisfies statement
3 (or equivalently, it is trivial). We only need to show that, for
any elementary rule, if we start with trivial EIIs, then the new EIIs
derived are still trivial. It is clear that (Sha) gives a trivial
EII, and (Cone) and (Abs) give trivial EIIs if we start with trivial
EIIs. For (SubV), if $\mathbf{A}\stackrel{EII}{\to}\mathbf{B}$ is
trivial, then there exists $\mathbf{T}\in\{0,1\}^{l\times n}$ and
a matrix $\mathbf{D}$ with nonnegative entries such that $\mathbf{B}\mathrm{sub}(\mathbf{I}_{n};\mathbf{T})=\mathbf{D}\left[\begin{array}{c}
\mathbf{A}\\
\boldsymbol{\Gamma}_{n}
\end{array}\right]$. Fix any $\mathbf{S}\in\{0,1\}^{n\times n'}$. By the multiplicativity
of $\mathrm{sub}$ (Proposition \ref{prop:sub_mul}), we have
\begin{align*}
 & \mathbf{B}\mathrm{sub}\left(\left[\begin{array}{c|c}
\mathbf{S} & \mathbf{0}^{n\times l}\\
\hline \mathbf{0}^{l\times n'} & \mathbf{I}_{l}
\end{array}\right]\right)\mathrm{sub}(\mathbf{I}_{n'};\mathbf{T}\mathbf{S})\\
 & =\mathbf{B}\mathrm{sub}(\mathbf{S};\mathbf{T}\mathbf{S})\\
 & =\mathbf{B}\mathrm{sub}(\mathbf{I}_{n};\mathbf{T})\mathrm{sub}(\mathbf{S})\\
 & =\mathbf{D}\left[\begin{array}{c}
\mathbf{A}\mathrm{sub}(\mathbf{S})\\
\boldsymbol{\Gamma}_{n}\mathrm{sub}(\mathbf{S})
\end{array}\right].
\end{align*}
Note that each row of $\boldsymbol{\Gamma}_{n}\mathrm{sub}(\mathbf{S})$
corresponds to a Shannon-type inequality, and hence is a conic combination
of rows of $\boldsymbol{\Gamma}_{n'}$. Therefore, the EII
\begin{align*}
 & \mathbf{A}\mathrm{sub}(\mathbf{S})\stackrel{EII}{\to}\mathbf{B}\mathrm{sub}\left(\left[\begin{array}{c|c}
\mathbf{S} & \mathbf{0}^{n\times l}\\
\hline \mathbf{0}^{l\times n'} & \mathbf{I}_{l}
\end{array}\right]\right)
\end{align*}
also satisfies statement 3.

For (SubA), let $\mathbf{S}\in\{0,1\}^{l\times(n+l')}$, and assume
$\mathbf{A}\stackrel{EII}{\to}\mathbf{B}\mathrm{sub}\left(\left[\begin{array}{c}
\mathbf{I}_{n}\,|\,\mathbf{0}^{n\times l'}\\
\hline \mathbf{S}
\end{array}\right]\right)$ is trivial. There exists $\mathbf{T}\in\{0,1\}^{l'\times n}$ and
a matrix $\mathbf{D}$ with nonnegative entries such that 
\begin{align*}
\mathbf{B}\mathrm{sub}\left(\left[\begin{array}{c}
\mathbf{I}_{n}\,|\,\mathbf{0}^{n\times l'}\\
\hline \mathbf{S}
\end{array}\right]\right)\mathrm{sub}(\mathbf{I}_{n};\mathbf{T}) & =\mathbf{D}\left[\begin{array}{c}
\mathbf{A}\\
\boldsymbol{\Gamma}_{n}
\end{array}\right].
\end{align*}
By the multiplicativity of $\mathrm{sub}$,
\[
\mathbf{B}\mathrm{sub}\left(\left[\begin{array}{c}
\mathbf{I}_{n}\,|\,\mathbf{0}^{n\times l'}\\
\hline \mathbf{S}
\end{array}\right]\right)\mathrm{sub}(\mathbf{I}_{n};\mathbf{T})=\mathbf{B}\mathrm{sub}(\mathbf{I}_{n};\mathbf{U}).
\]
where $\mathbf{U}=\mathbf{S}\left[\begin{array}{c}
\mathbf{I}_{n}\\
\mathbf{T}
\end{array}\right]\in\mathbb{R}_{\ge0}^{l\times n}$. Therefore, $\mathbf{A}\stackrel{EII}{\to}\mathbf{B}$ also satisfies
statement 3.

For (Tran), assume $\mathbf{A}\stackrel{EII}{\to}\mathbf{B}$ and
$\mathbf{B}\stackrel{EII}{\to}\mathbf{C}$ are trivial for $\mathbf{A}\in\mathbb{R}^{m_{\mathbf{A}}\times(2^{n}-1)}$,
$\mathbf{B}\in\mathbb{R}^{m_{\mathbf{B}}\times(2^{n+l}-1)}$, $\mathbf{C}\in\mathbb{R}^{m_{\mathbf{C}}\times(2^{n+l+k}-1)}$.
There exist $\mathbf{S}\in\{0,1\}^{l\times n}$, $\mathbf{T}\in\{0,1\}^{k\times(n+l)}$
and matrices $\mathbf{D},\mathbf{E}$ with nonnegative entries such
that 
\[
\mathbf{B}\mathrm{sub}(\mathbf{I}_{n};\mathbf{S})=\mathbf{D}\left[\begin{array}{c}
\mathbf{A}\\
\boldsymbol{\Gamma}_{n}
\end{array}\right],
\]
\[
\mathbf{C}\mathrm{sub}(\mathbf{I}_{n+l};\mathbf{T})=\mathbf{E}\left[\begin{array}{c}
\mathbf{B}\\
\boldsymbol{\Gamma}_{n+l}
\end{array}\right].
\]
Let $\mathbf{U}=\mathbf{T}\left[\begin{array}{c}
\mathbf{I}_{n}\\
\mathbf{S}
\end{array}\right]\in\mathbb{R}_{\ge0}^{k\times n}$. We have
\begin{align*}
 & \mathbf{C}\mathrm{sub}(\mathbf{I}_{n};\mathbf{S};\mathbf{U})\\
 & =\mathbf{C}\mathrm{sub}(\mathbf{I}_{n+l};\mathbf{T})\mathrm{sub}(\mathbf{I}_{n};\mathbf{S})\\
 & =\mathbf{E}\left[\begin{array}{c}
\mathbf{B}\mathrm{sub}(\mathbf{I}_{n};\mathbf{S})\\
\boldsymbol{\Gamma}_{n+l}\mathrm{sub}(\mathbf{I}_{n};\mathbf{S})
\end{array}\right]\\
 & =\mathbf{E}\left[\begin{array}{c}
\mathbf{D}\mathbf{A}\\
\mathbf{D}\boldsymbol{\Gamma}_{n}\\
\boldsymbol{\Gamma}_{n+l}\mathrm{sub}(\mathbf{I}_{n};\mathbf{S})
\end{array}\right].
\end{align*}
Note that each row of $\boldsymbol{\Gamma}_{n+l}\mathrm{sub}(\mathbf{I}_{n};\mathbf{S})$
corresponds to a Shannon-type inequality, and hence is a conic combination
of rows of $\boldsymbol{\Gamma}_{n}$. Therefore, the EII $\mathbf{A}\stackrel{EII}{\to}\mathbf{C}$
also satisfies statement 3.

\[
\]

\bibliographystyle{IEEEtran}
\bibliography{ref}

\end{document}